\documentclass[aps,pre,reprint,twocolumn,showpacs,floatfix]{revtex4-1}

\usepackage{graphicx}
\usepackage{amsmath}
\usepackage{amssymb}
\usepackage{epsf}
\usepackage{color}

\begin{document}

\title{Information theoretical study of cross-talk mediated signal 
transduction in MAPK pathways}

\author{Alok Kumar Maity}
\email{amaity@ucla.edu}
\affiliation{Department of Chemistry, University of Calcutta, 
92 A P C Road, Kolkata 700009, India}
\altaffiliation{Present address: 
Department of Chemistry and Biochemistry, University of California, 
Los Angeles, Los Angeles, CA 90095, United States of America}

\author{Pinaki Chaudhury}
\email{pinakc@rediffmail.com}
\affiliation{Department of Chemistry, University of Calcutta, 
92 A P C Road, Kolkata 700009, India}

\author{Suman K Banik}
\email{skbanik@jcbose.ac.in}
\affiliation{Department of Chemistry, Bose Institute, 
93/1 A P C Road, Kolkata 700009, India}

\begin{abstract}
Biochemical networks related to similar functional pathways are often correlated 
due to cross-talk among the homologous proteins in the different networks. 
Using a stochastic framework, we address the functional significance of the 
cross-talk between two pathways. Our theoretical analysis on generic MAPK 
pathways reveals cross-talk is responsible for developing coordinated fluctuations 
between the pathways. The extent of correlation evaluated in terms of the information 
theoretic measure provides directionality to net information propagation. Stochastic 
time series and scattered plot suggest that the cross-talk generates synchronization 
within a cell as well as in a cellular population. Depending on the number of input 
and output, we identify signal integration and signal bifurcation motif that arise due 
to inter-pathway connectivity in the composite network. Analysis using partial 
information decomposition quantifies the net synergy in the information propagation 
through these branched pathways.   
\end{abstract}


\maketitle


\section{Introduction}

The decision making processes at the cellular level are initiated by some 
specialized signaling networks \cite{Tyson2003,Alon2006}. These networks 
play a pivotal role in making robust and precise cellular response towards
endogenic and exogenic perturbations. In addition, the process of decision 
making resolves cellular fate as well as survival strategies in diverse peripheral 
conditions. Although both prokaryotic and eukaryotic cells are comprised of 
several common signaling networks, few signaling networks are incorporated 
mostly in the eukaryotes due to the evolutionary prospect \cite{Lyons2012}.  
One such signaling network is the mitogen-activated protein kinase (MAPK) 
pathway that plays the central role to attune with extra-cellular signal in 
eukaryotic cells
\cite{Bardwell2006,Ferrell1996,Huang1996,Saito2010,Voliotis2014}. 
Although different MAPK pathways with diverse inputs and outputs belong 
to a higher living species, they are sometimes interconnected through 
overlapping sets of signaling components. Depending on the interconnections, 
MAPK pathways can be classified into different groups that use one or more 
than one common signaling components. Moreover, as a result of cross-interaction, 
a single regulon regulates multiple targets in addition to its own target. Such 
type of signal association is defined as cross-talk. Cross-coupling in the 
signaling network can modify the functionality of a network topology and 
can subsume errors compared to the uncoupled one. Cross-interactions 
have been identified not only in eukaryotes but also in prokaryotes, as 
observed in the bacterial two-component system 
\cite{Laub2007,Posas1996,Rowland2014,Siryaporn2010,Trach1993}. 
In the eukaryotic system, cross-talk has been identified in numerous 
situations\cite{Gustin1998,Kunkel2002,Oeckinghaus2011}. Furthermore, 
cross-talk and several of its variance have also been identified at different 
stages of gene regulation 
\cite{Iborra2004,Mather2013, Mauri2014,Riba2014,Tsimring2014}.

Since cross-talk is observed in a broad range of biological processes, one 
may interrogate the functional utility of such network coordination. The 
cross-coupling mechanism is conveyed through generations despite the 
continuous development of a large number of evolutionary descent with
modifications in the cellular interactions within the surviving trait. This 
character indicates that crosstalk might have a definite functional potential 
to build up synchronized cellular regulations by spending the storage energy. 
If this is true, how a cell balances the trade-off between network association 
and potential cost? Few comprehensive experiments on the network connectivity 
suggest that networks of a well delineated cluster are correlated with each other 
but are uncorrelated to the rest of the network \cite{Saito2010}. Synchronization 
is necessary to attain natural activity but needs to maintain a threshold value. 
Otherwise too much synchronization may lead to physiological disorder like 
epilepsy \cite{Palus2001}. Inter-pathway cross-talk becomes prominent due to 
limitation of common resources, defined as overloaded condition. However, 
cross-talk effect becomes faint in the underloaded condition, where level of 
available resources is satisfactory 
\cite{Cookson2011,Komili2008,Mather2013,Mauri2014,Tsimring2014}. A key 
source of survival strategy under diverse environmental conditions is the 
generation of fluctuations which induces non-genetic variability in a cellular 
population. In such a situation, cells readjust to cope with the limited resources 
by introducing cross-correlation among a set of genes and thus implementing 
a successful bet-hedging program \cite{Tsimring2014}. Cross-talk also facilitates 
synchronization in different organs such as cardio-respiratory interaction, brain 
and tissues \cite{Palus2001}.

To address the functionality of cross-talk, we undertake a representative cluster 
of networks with overlapping sets of regulatory components. To be specific, we 
focus on MAPK pathway, a well-studied eukaryotic signaling machinery, conserved 
with three kinase cascades. In \textit{S. cerevisiae}, five MAPK signaling pathways 
are present among which only three (pheromone response, filamentous growth 
response and osmostress adaptation) use a common kinase protein Ste11 
\cite{Bardwell2006,Saito2010}. In fact, pheromone response and filamentous 
growth pathways also use the same kinase Ste7. Pheromone MAPK cascade 
(Ste11 $\rightarrow$ Ste7 $\rightarrow$ Fus3) is activated by mating pheromone. 
Under low nutrient condition, filamentous growth MAPK cascade (Ste11 
$\rightarrow$ Ste7 $\rightarrow$ Kss1)gets activated whereas high external 
osmolarity activates the osmoadaption cascade (Ste11 $\rightarrow$ Pbs2 
$\rightarrow$ Hog1) \cite{Bardwell2006,Saito2010}. Due to intercascade 
correlation among the three signaling pathways, one pathway can be activated 
by the signal of another pathway in absence of its own signal. Several experimental 
results suggest that such cross-talk is filtered out by cross-pathway inhibition, 
kinetic insulation and formation of scaffold protein 
\cite{Bardwell2006,Bardwell2007,Komarova2005,McClean2007,Saito2010,Ubersax2007}. 
Although activation through inter-pathway cross-talk and cross-pathway 
inhibition compensates each other, information is exchanged among the 
pathways during these interactions. This leads to obvious queries (i) is it 
possible for an individual signaling pathway to convey its input signal reliably 
downstream without experiencing any influence from the other pathways signal? 
(ii) Since the inter-pathway connectivity is known not to allow the uniqueness of 
transduced signals - what are the physiological advantages of cross-association? 
(iii) Is there any participation of pathway output in the cooperative regulation 
of a downstream target in a synchronized manner? (iv) How is it possible for 
correlated pathways to keep up static as well as dynamic synchronization in 
a single cell environment that is prevalently stochastic in nature? (v) Does 
this association have any capability to control the cell-to-cell variability?

In the present manuscript, we study generic \textit{S. cerevisiae} MAPK 
pathways to address the potential functionality of inter-pathway cross-talk 
within a stochastic framework. We consider two equivalent interacting 
MAPK pathways, each one consisting of a linear chain of three MAPK 
cascade proteins \cite{Bardwell2006,Saito2010}. Both pathways get 
stimulated by their corresponding external signals propagating downstream 
through phosphorylation (activation) and dephosphorylation (deactivation) 
of the cascade proteins. In addition, due to cross-talk, phosphorylation of the 
intermediate components of the two pathways is influenced by the activated 
kinase of the other pathway along with the cognate one. As the population 
of each cascade protein is not sufficiently high within a single cell and 
experiences a fluctuating environment, we express all associated chemical 
reactions in terms of stochastic differential equation. We solve the coupled 
set of nonlinear Langevin equations using linear noise approximation 
\cite{Kampen2011,Gardiner2009} and 
calculate the auto variance of each and every kinase and covariance 
between two different kinases (see Sec.~II and Appendix). 
Recent theoretical 
development \cite{Grima2015} shows that linear noise approximation is 
not only limited for high copy number but also exact up to second moments 
of any chemical species participating in a second-order reaction. The fluctuations 
associated with at least one of the species participating in each of the 
second-order reaction are Poissonian and uncorrelated with the 
fluctuations of other species. In addition, linear noise approximation 
remains valid for faster activation and deactivation (or synthesis and 
degradation) rates of the corresponding components compared to the 
coarse-grained (steady state) time scale \cite{Elf2003,Gardiner2009,
Maity2014,Maity2015,Paulsson2004,Paulsson2005,Kampen2011,
Grima2011,Thomas2013,ThomasP2013,Grima2015}. To classify the signal 
transduction efficacy through two pathways 
in the presence of cross-association, we quantify two as well as three variable 
mutual information. Distributions of all kinase proteins are approximately 
considered Gaussian, allowing us to adopt a reduced expression of mutual 
information \cite{Cover2012,Shannon1948}. The reduced equation mainly 
depends on the auto variance and the covariance of the corresponding 
kinase. We validate our analytical calculation by exact stochastic simulation 
\cite{Gillespie1976}. In the first subsection, we quantify two variable mutual 
information under the influence of cross-talk parameter. We also investigate 
the mutual information between two non-cognate kinases and find causality 
of this coordination. Since causality leads to synchronization 
\cite{Granger1969,Palus2001}. it is important to measure causality relation 
between the pathways, i.e., who regulates whom and to which extent. If
both pathways interact with each other and transduce information of the 
corresponding input signal with different degrees, then it is very difficult to 
characterize the magnitude and direction of signal propagation. To overcome 
such difficulty, we define a new measure, net information transduction,
using the expressions of two cross mutual information, which satisfactorily 
quantifies the amount of net signal propagation. We also verify inter-pathway 
synchronization with the help of coordinated fluctuations of stochastic 
trajectories of two parallel kinases. This result implies how two kinases are
synchronized within a cell. To understand this phenomenon further, we 
investigate how much association is developed among the steady state 
population of these kinases from cell-to-cell. In the second subsection, we 
quantify three variable mutual information when both the channels of 
information flow work separately. Applying the theory of partial information 
decomposition \cite{Barrett2015,Schneidman2003},  we quantify the net synergy. 
We observe the sign of net synergy value changes depending on the signal 
integration as well as signal bifurcation and is mainly controlled by pathway 
architecture.


\section{The Model}

In Fig.~1 we show a schematic diagram of two interacting parallel MAPK pathways (named as X and
Y). Each MAPK pathway consists of three kinase components, i.e., $x_1$, $x_2$, $x_3$ (X pathway) 
and $y_1$, $y_2$, $y_3$ (Y pathway) \cite{Heinrich2002,Huang1996,McClean2007,
Suderman2013,Tanase-Nicola2006}. $x_i$ and $x_{pi}$ represent dephosphorylated and phosphorylated 
form of a kinase protein, respectively, and the same applies to  $y_j$ and $y_{pj}$ (here $i,j=1,2,3$). 
The first cascade protein of a MAPK pathway gets phosphorylated with an exposure to the external 
stimulus. While phosphorylated, it positively regulates the phosphorylation of its own downstream 
kinase along with the kinase of the other pathway. The phosphorylated intermediate kinase 
regulates phosphorylation of the last kinase. To maintain the pool of phosphorylated kinase 
within a cell, a dephosphorylation process is in action with the help of phosphatase molecules. 
The cross-pathway interactions between two parallel MAPK pathways are denoted by the dashed 
lines in Fig.~1 along with the cross-interaction rate parameters ${\varepsilon}_1$ and 
${\varepsilon}_2$. $S_x$ and $S_y$ are the two extra-cellular signals acting on the X and 
Y pathway, respectively.

Both pathways get causally correlated through cross-interactions, and a cross-talk develops as a
consequence. Causal relationships are frequently examined in various circumstances that are
subjected to stochastic fluctuations \cite{Barrett2015,Granger1969,Palus2001,Schneidman2003}. 
In the present manuscript, we quantify the causal relationship in terms of mutual information. 
Here, the two cross-interaction parameters ${\varepsilon}_1$ and ${\varepsilon}_2$ play a 
significant role in establishing different levels of cross-talk. The parameter ${\varepsilon}_1$ 
controls information flow from X to Y pathway ($x_{p1} \rightarrow y_{p2}$), but the parameter 
${\varepsilon}_2$ is responsible for Y to X pathway ($y_{p1} \rightarrow x_{p2}$) information flow. 
In this connection, it is important to mention that during mating process, both pheromone and 
filamentous growth pathways are activated to a roughly equal extent, whereas during
invasive growth process, only filamentous growth pathway is activated \cite{Bardwell2007}. 
These observations corroborate with our model development. In our calculation, we only consider 
the post-translationally modified forms of all MAPK proteins. Thus, in the model, the total 
population of a MAPK protein is the sum of the phosphorylated and the unphosphorylated form 
of the protein and is considered to be a constant ($(x_i+x_{pi})=
x_{Ti}=(y_j+y_{pj})=y_{Tj}=$ constant, here $i=j$). In addition, we consider a physiologically relevant 
parameter set for our calculation \cite{Bardwell1996,Ferrell1996,Voliotis2014}.


\begin{figure}[!t]
\includegraphics[width=1.0\columnwidth,angle=0]{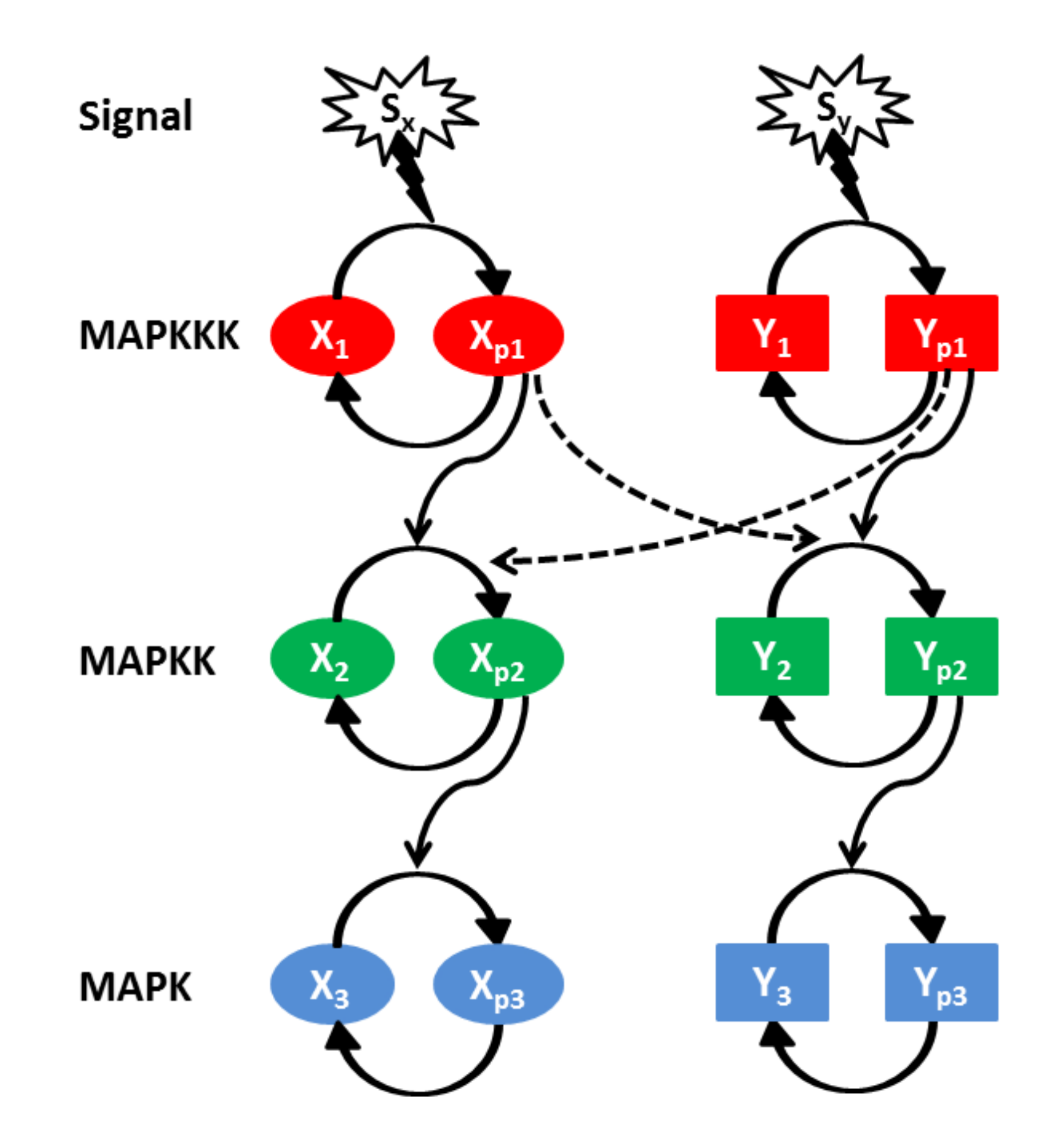}
\caption{(color online)
Schematic diagram of two parallel MAPK (equivalent and identical) signaling pathways (X and Y).
Each pathway consists of three successively connected cascade kinases, MAPKKK (red), MAPKK 
(green) and MAPK (blue). The first activated kinase facilitates the activation of the second one and
then the second kinase regulates the activation of the last one. Both signaling pathways are exposed
to two different signals ($S_x$ and $S_y$). Cross-talk is developed due to inter-pathway 
interactions. $\varepsilon_{1}$ and $\varepsilon_{2}$ are the cross-interaction parameters and 
the directionality of these interactions are $x_{p1} \rightarrow y_{p2}$ and 
$y_{p1} \rightarrow x_{p2}$, respectively.
}
\label{fig1}
\end{figure}

\subsection{Two variable mutual information}

Adopting Shannon's information theory \cite{Cover2012,Shannon1948}, we have calculated 
two variable mutual information between two phosphorylated kinases,
\begin{equation}
\label{eq1}
{\cal I} (x_{pi}; y_{pj}) = \sum_{x_{pi}} \sum_{y_{pj}} p(x_{pi}, y_{pj}) \log_2 
\left [\frac {p(x_{pi}, y_{pj})}{p(x_{pi})p(y_{pj})} \right ].
\end{equation}  
A generalised index $x_{pi}$ and $y_{pj}$ have been considered to represent the 
copy number of two different phosphorylated kinases. Similarly, $p(x_{pi})$ and $p(y_{pj})$ 
are the marginal and $p(x_{pi}, y_{pj})$ is the joint probability distributions associated 
with the corresponding kinases. For the calculation of mutual information between two kinases of 
X signaling pathway, we have replaced $y_{pj}$ by $x_{pi}$ (where $i \neq j$)and the reverse 
replacement has been followed for Y signaling pathway. For the estimation of mutual
information between two equivalent kinases ($x_{pi}$ and $y_{pj}$) of the respective pathways, 
we have used the same formula for $i = j$ condition. Mutual information can also be written in 
the form of the entropy function. Hence, Eq.~(\ref{eq1}) can be redefined as 
\begin{equation}     
\label{eq2}
{\cal I} (x_{pi}; y_{pj}) = H(x_{pi}) + H(y_{pj}) - H(x_{pi}, y_{pj}).
\end{equation}
Here, $H(x_{pi})$ and $H(y_{pj})$ are individual and $H(x_{pi}, y_{pj})$ is total entropy of the 
respective kinases. In the present study, both probability distribution functions (marginal 
as well as joint) are approximately considered to be Gaussian. Thus, using Gaussian channel 
approximation \cite{Barrett2015,Cover2012,Shannon1948}, Eq.~(\ref{eq2}) takes
the reduced form
\begin{equation}
\label{eq3}
{\cal I} (x_{pi}; y_{pj}) = \frac{1}{2} \log_2 \left [ \frac
{\sigma_{x_{pi}}^2 \sigma_{y_{pi}}^2} {\sigma_{x_{pi}}^2 \sigma_{y_{pi}}^2 - \sigma_{x_{pi} y_{pi}}^4} \right] ,
\end{equation}  
where $\sigma_{x_{pi}}^2$ and $\sigma_{y_{pj}}^2$ are auto variances and 
$\sigma_{x_{pi} y_{pj}}^2$ is covariance of the corresponding kinases (for 
detailed calculation see Appendix). 
At this point it is important to mention that no prior knowledge is required about 
the nature of probability distribution function for evaluating mutual information 
using Eq.~(\ref{eq1}). For exact or approximate Gaussian distribution, one can 
reduce Eq.~(\ref{eq1}) to Eq.~(\ref{eq3}) applying Gaussian channel approximation. 
However, for systems with non-Gaussian distribution, one can still use Eq.~(\ref{eq1}) 
with proper analytical expressions of probability distribution functions that 
may contribute expressions of higher moments in Eq.~(\ref{eq3}).

In the present work, all expressions of 
two variable mutual information are calculated using Eq.~(\ref{eq3}). The 
analytical results are then validated by evaluating probability distribution 
functions (Eq.~(\ref{eq1})) using exact numerical simulation \cite{Gillespie1976}. 
In our numerical simulation we have used $10^7$ trajectories and smallest bin 
size of $1$. The two variable mutual information value is bounded with in a 
scale $0 \leq {\cal I} (x_{pi}; y_{pj}) \leq \min(H(x_{pi}), H(y_{pj}))$. To quantify 
the association between two equivalent kinases, we have used Pearson's 
correlation coefficient ($\rho_{ij}, i=j$) \cite{Whitlock2009}
\begin{equation}  
\label{eq4}
\rho_{ij} = \frac {\sigma_{x_{pi} y_{pj}}^2}{\sigma_{x_{pi}} \sigma_{y_{pj}}}.
\end{equation}  .

\subsection{Three variable mutual information}

The three variable mutual information are calculated for both signal integration 
and signal bifurcation motif. In the first motif, two phosphorylated input kinases 
interact with one output kinase. Hence the complete description of mutual 
information is given by      
\begin{eqnarray}
\label{eq5}
{\cal I} (x_{p1}, y_{p1}; x_{p2}) & = & 
\sum_{x_{p1}, y_{p1}} \sum_{x_{p2}} p(x_{p1}, y_{p1}, x_{p2}) \nonumber \\
&& \times \log_2 
\left [\frac {p(x_{p1}, y_{p1}, x_{p2})}{p(x_{p1}, y_{p1})p(x_{p2})} \right],
\end{eqnarray}

\noindent
where $p(x_{p1}, y_{p1}, x_{p2})$ and $p(x_{p1}, y_{p1})$ are the joint distribution 
functions of the corresponding components. On the other hand, $p(x_{p2})$ is the 
marginal distribution of phosphorylated $x_2$ kinase. One can also write 
Eq.~(\ref{eq5}) in terms of the respective entropy
\begin{equation} 
\label{eq6}
{\cal I} (x_{p1}, y_{p1}; x_{p2}) = H(x_{p1}, y_{p1}) + H(x_{p2}) - H(x_{p1}, y_{p1}, x_{p2}).
\end{equation}

\noindent
Similarly, using Gaussian approximation \cite{Barrett2015,Cover2012,Shannon1948},
one can reduce Eq.~(\ref{eq6}) into the following form
\begin{equation}
\label{eq7}
{\cal I} (x_{p1}, y_{p1}; x_{p2}) = \frac{1}{2} \log_2 
\left [\frac {\sigma_{x_{p2}}^2 (\sigma_{x_{p1}}^2 \sigma_{y_{p1}}^2 - \sigma_{x_{p1} y_{p1}}^4)}
{| \Delta_1 |} \right ],
\end{equation}

\noindent
with
\begin{eqnarray}
| \Delta_1 | &=&
\left ( 
\begin{array}{ccc}
\sigma_{x_{p1}}^2 & \sigma_{x_{p1} y_{p1}}^2 & \sigma_{x_{p1} x_{p2}}^2 \\
\sigma_{y_{p1} x_{p1}}^2 & \sigma_{y_{p1}}^2 & \sigma_{y_{p1} x_{p2}}^2 \\
\sigma_{x_{p2} x_{p1}}^2 & \sigma_{x_{p2} y_{p1}}^2 & \sigma_{x_{p2}}^2 \nonumber
\end{array}
\right ) .
\end{eqnarray}

\noindent
Here, the magnitude of three variable mutual information is bounded within a scale 
$0 \leq {\cal I} (x_{p1}, y_{p1}; x_{p2}) \leq \min(H(x_{p1}, y_{p1}), H(x_{p2}))$. 
Using partial information decomposition formalism, the three variable mutual 
information can be decomposed into two parts \cite{Barrett2015,Schneidman2003}. 
As a result, the net synergy expression becomes
\begin{widetext}
\begin{eqnarray}
\label{eq8}
\Delta {\cal I} (x_{p1}, y_{p1}; x_{p2}) 
& = &  {\cal I} (x_{p1}, y_{p1}; x_{p2}) - {\cal I} (x_{p1}; x_{p2}) 
- {\cal I} (y_{p1}; x_{p2}) \nonumber \\
& = &  
\frac {1}{2} \log_2 \left [\frac 
{(\sigma_{x_{p1}}^2 \sigma_{y_{p1}}^2 - \sigma_{x_{p1} y_{p1}}^4) (\sigma_{x_{p1}}^2 \sigma_{x_{p2}}^2 - \sigma_{x_{p1} x_{p2}}^4) (\sigma_{y_{p1}}^2 \sigma_{x_{p2}}^2 - \sigma_{y_{p1} x_{p2}}^4)}
{| \Delta_1 | \sigma_{x_{p1}}^2 \sigma_{y_{p1}}^2 \sigma_{x_{p2}}^2} \right ] .
\end{eqnarray}  
\end{widetext}

\noindent 
Furthermore, one can calculate mutual information for the signal bifurcating motif 
with the help of associated distribution functions
\begin{eqnarray}
\label{eq9}
{\cal I} (x_{p1}; x_{p2}, y_{p2}) & = & 
\sum_{x_{p1}} \sum_{x_{p2}, y_{p2}} p(x_{p1}, x_{p2}, y_{p2}) \nonumber \\
&& \times \log_2 
\left [\frac {p(x_{p1}, x_{p2}, y_{p2})}{p(x_{p1})p(x_{p2}, y_{p2})} \right].
\end{eqnarray}

\noindent
and the entropy representation of Eq.~(\ref{eq9}) is
\begin{equation} 
\label{eq10}
{\cal I} (x_{p1}; x_{p2}, y_{p2}) = H(x_{p1}) + H(x_{p2}, y_{p2}) - H(x_{p1}, x_{p2}, y_{p2}) .
\end{equation}

\noindent
Using Gaussian approximation \cite{Barrett2015,Cover2012,Shannon1948} 
Eq.~(\ref{eq10}) becomes
\begin{equation}
\label{eq11}
{\cal I} (x_{p1}; x_{p2}, y_{p2}) = \frac{1}{2} \log_2 
\left [\frac {\sigma_{x_{p1}}^2 (\sigma_{x_{p2}}^2 \sigma_{y_{p2}}^2 - \sigma_{x_{p2} y_{p2}}^4)}
{| \Delta_2 |} \right ],
\end{equation}

\noindent
with  
\begin{eqnarray}
| \Delta_2 | &=&
\left ( 
\begin{array}{ccc}
\sigma_{x_{p1}}^2 & \sigma_{x_{p1} x_{p2}}^2 & \sigma_{x_{p1} y_{p2}}^2 \\
\sigma_{x_{p2} x_{p1}}^2 & \sigma_{x_{p2}}^2 & \sigma_{x_{p2} y_{p2}}^2 \\
\sigma_{y_{p2} x_{p1}}^2 & \sigma_{y_{p2} x_{p2}}^2 & \sigma_{y_{p2}}^2 \nonumber
\end{array}
\right ) .
\end{eqnarray}

\noindent
In addition, the three variable mutual information value is bounded within a range 
$0 \leq {\cal I} (x_{p1}; x_{p2}, y_{p2}) \leq \min(H(x_{p1})$, $H(x_{p2}, y_{p2}))$. 
In this case, one can also use the theory of partial information decomposition to 
decompose the three variable mutual information into two parts and calculate 
the net synergy \cite{Schneidman2003}
\begin{widetext}
\begin{eqnarray}
\label{eq12}
\Delta {\cal I} (x_{p1}; x_{p2}, y_{p2}) & = & {\cal I} (x_{p1}; x_{p2}, y_{p2}) - {\cal I} (x_{p1}; x_{p2}) 
- {\cal I} (x_{p1}; x_{p2}) \nonumber \\
& = & 
\frac {1}{2} \log_2 \left [\frac 
{(\sigma_{x_{p2}}^2 \sigma_{y_{p2}}^2 - \sigma_{x_{p2} y_{p2}}^4) (\sigma_{x_{p1}}^2 \sigma_{x_{p2}}^2 - \sigma_{x_{p1} x_{p2}}^4) (\sigma_{x_{p1}}^2 \sigma_{y_{p2}}^2 - \sigma_{x_{p1} y_{p2}}^4)}
{| \Delta_2 | \sigma_{x_{p1}}^2 \sigma_{x_{p2}}^2 \sigma_{y_{p2}}^2} \right ] .
\end{eqnarray}  
\end{widetext}

\noindent
For analytical calculation, we have adopted 
Eqs.~(\ref{eq3}, \ref{eq4}, \ref{eq7}, \ref{eq8}, \ref{eq11} and \ref{eq12}) which 
contain only auto variance and covariance expressions, whereas we adopt 
numerical simulation for evaluation of the expressions given in Eqs.~(\ref{eq1}, 
\ref{eq5} and \ref{eq9}).
At this point it is important to mention that we validate our analytical calculation 
by exact stochastic simulation, commonly known as stochastic simulation 
algorithm or Gillespie algorithm \cite{Gillespie1976}. The validation signifies how 
much closer the system dynamics with the Gaussian statistics. In the following 
section, corroboration of analytical and simulation results indicate a valid 
consideration of Gaussian approximation.


\begin{table*}[!ht]
\caption{
Reactions and corresponding parameter values for the MAPK network motif of 
\textit{S. cerevisiae} \cite{Bardwell1996,Ferrell1996,Voliotis2014}, related to faster 
relaxation rate.\footnote{Other Parameters are
$s_x = s_y = 10$ molecules/cell,
$x_{T1} = x_1 + x_{p1} = 250$ molecules/cell,
$x_{T2} = x_2 + x_{p2} = 1700$ molecules/cell,
$x_{T3} = x_3 + x_{p3} = 5000$ molecules/cell,
$y_{T1} = y_1 + y_{p1} = 250$ molecules/cell,
$y_{T2} = y_2 + y_{p2} = 1700$ molecules/cell and
$y_{T3} = y_3 + y_{p3} = 5000$ molecules/cell.}
The kinetic schemes adopted in the present work follows the model
of Heinrich et al \cite{Heinrich2002}.
}
\begin{ruledtabular}
\begin{tabular}{lllll} 
Description & Reaction &  & Propensity function & Rate constant \\
\hline
Activation of $x_1$  & $ x_1 + s_x \stackrel{k_x}{\longrightarrow} x_{p1} + s_x $
& & $ k_x s_x x_1 $
& $ k_x =10^{-4} $ molecules$^{-1}$ s$^{-1} $ \\
Deactivation of $x_{p1}$  &  $x_{p1} \stackrel{\alpha_{1}}{\longrightarrow} x_1$ 
& & $ \alpha_{1} x_{p1} $
& $ \alpha_{1} = 0.01$ s$^{-1} $ \\
Activation of $y_1$  & $ y_1 + s_y \stackrel{k_y}{\longrightarrow} y_{p1} + s_y $
& & $ k_y s_y y_1 $
& $ k_y =10^{-4}$  molecules$^{-1}$ s$^{-1} $ \\
Deactivation of $y_{p1}$  &  $y_{p1} \stackrel{\beta_{1}}{\longrightarrow} y_1$ 
& & $ \beta_{1} y_{p1} $
& $ \beta_{1} = 0.01$ s$^{-1} $ \\
Activation of $x_2$  & $ x_2 + x_{p1} \stackrel{k_{12x}}{\longrightarrow} x_{p2} + x_{p1} $
& & $ k_{12x} x_{p1} x_2 $
& $ k_{12x} =10^{-4}$  molecules$^{-1}$ s$^{-1} $ \\
Activation of $x_2$  & $ x_2 + y_{p1} \stackrel{\varepsilon_{2}}{\longrightarrow} x_{p2} + y_{p1} $
& & $ \varepsilon_{2} y_{p1} x_2 $
& $ \varepsilon_{2} = (0-1)\times 10^{-4}$  molecules$^{-1}$ s$^{-1} $ \\
Deactivation of $x_{p2}$  &  $x_{p2} \stackrel{\alpha_{2}}{\longrightarrow} x_2$ 
& & $ \alpha_{2} x_{p2} $
& $ \alpha_{2} = 0.05$ s$^{-1} $ \\
Activation of $y_2$  & $ y_2 + y_{p1} \stackrel{k_{12y}}{\longrightarrow} y_{p2} + y_{p1} $
& & $ k_{12y} y_{p1} y_2 $
& $ k_{12y} =10^{-4}$  molecules$^{-1}$ s$^{-1} $ \\
Activation of $y_2$  & $ y_2 + x_{p1} \stackrel{\varepsilon_{1}}{\longrightarrow} y_{p2} + x_{p1} $
& & $ \varepsilon_{1} x_{p1} y_2 $
& $ \varepsilon_{1} = (0-1)\times 10^{-4}$  molecules$^{-1}$ s$^{-1} $ \\
Deactivation of $y_{p2}$  &  $y_{p2} \stackrel{\beta_{2}}{\longrightarrow} y_2$ 
& & $ \beta_{2} y_{p2} $
& $ \beta_{2} = 0.05$ s$^{-1} $ \\
Activation of $x_3$  & $ x_3 + x_{p2} \stackrel{k_{23x}}{\longrightarrow} x_{p3} + x_{p2} $
& & $ k_{23x} x_{p2} x_3 $
& $ k_{23x} = 5 \times 10^{-5}$  molecules$^{-1}$ s$^{-1} $ \\
Deactivation of $x_{p3}$  &  $x_{p3} \stackrel{\alpha_{3}}{\longrightarrow} x_3$ 
& & $ \alpha_{3} x_{p3} $
& $ \alpha_{3} = 0.05$ s$^{-1} $ \\
Activation of $y_3$  & $ y_3 + y_{p2} \stackrel{k_{23y}}{\longrightarrow} y_{p3} + y_{p2} $
& & $ k_{23y} y_{p2} y_3 $
& $ k_{23y} = 5 \times 10^{-5}$  molecules$^{-1}$ s$^{-1} $ \\
Deactivation of $y_{p3}$  &  $y_{p3} \stackrel{\beta_{3}}{\longrightarrow} y_3$ 
& & $ \beta_{3} y_{p3} $
& $ \beta_{3} = 0.05$ s$^{-1} $ \\
\end{tabular}
\end{ruledtabular}
\label{table1} 
\end{table*}


\begin{table*}[!ht]
\caption{
Reactions and corresponding parameter values for the MAPK network motif of 
\textit{S. cerevisiae} \cite{Bardwell1996,Ferrell1996,Voliotis2014}, related to 
slower relaxation rate.\footnote{Other Parameters are
$s_x = s_y = 10$ molecules/cell,
$x_{T1} = x_1 + x_{p1} = 250$ molecules/cell,
$x_{T2} = x_2 + x_{p2} = 1700$ molecules/cell,
$x_{T3} = x_3 + x_{p3} = 5000$ molecules/cell,
$y_{T1} = y_1 + y_{p1} = 250$ molecules/cell,
$y_{T2} = y_2 + y_{p2} = 1700$ molecules/cell and
$y_{T3} = y_3 + y_{p3} = 5000$ molecules/cell.}
The kinetic schemes adopted in the present work follows 
the model of Heinrich et al \cite{Heinrich2002}.
}
\begin{ruledtabular}
\begin{tabular}{lllll} 
Description & Reaction &  & Propensity function & Rate constant \\
\hline
Activation of $x_1$  & $ x_1 + s_x \stackrel{k_x}{\longrightarrow} x_{p1} + s_x $
& & $ k_x s_x x_1 $
& $ k_x =10^{-4}$  molecules$^{-1}$ s$^{-1} $ \\
Deactivation of $x_{p1}$  &  $x_{p1} \stackrel{\alpha_{1}}{\longrightarrow} x_1$ 
& & $ \alpha_{1} x_{p1} $
& $ \alpha_{1} = 0.01$ s$^{-1} $ \\
Activation of $y_1$  & $ y_1 + s_y \stackrel{k_y}{\longrightarrow} y_{p1} + s_y $
& & $ k_y s_y y_1 $
& $ k_y =10^{-4}$  molecules$^{-1}$ s$^{-1} $ \\
Deactivation of $y_{p1}$  &  $y_{p1} \stackrel{\beta_{1}}{\longrightarrow} y_1$ 
& & $ \beta_{1} y_{p1} $
& $ \beta_{1} = 0.01$ s$^{-1} $ \\
Activation of $x_2$  & $ x_2 + x_{p1} \stackrel{k_{12x}}{\longrightarrow} x_{p2} + x_{p1} $
& & $ k_{12x} x_{p1} x_2 $
& $ k_{12x} =10^{-4}$  molecules$^{-1}$ s$^{-1} $ \\
Activation of $x_2$  & $ x_2 + y_{p1} \stackrel{\varepsilon_{2}}{\longrightarrow} x_{p2} + y_{p1} $
& & $ \varepsilon_{2} y_{p1} x_2 $
& $ \varepsilon_{2} = (0-1)\times 10^{-4}$  molecules$^{-1}$ s$^{-1} $ \\
Deactivation of $x_{p2}$  &  $x_{p2} \stackrel{\alpha_{2}}{\longrightarrow} x_2$ 
& & $ \alpha_{2} x_{p2} $
& $ \alpha_{2} = 0.01$ s$^{-1} $ \\
Activation of $y_2$  & $ y_2 + y_{p1} \stackrel{k_{12y}}{\longrightarrow} y_{p2} + y_{p1} $
& & $ k_{12y} y_{p1} y_2 $
& $ k_{12y} =10^{-4}$  molecules$^{-1}$ s$^{-1} $ \\
Activation of $y_2$  & $ y_2 + x_{p1} \stackrel{\varepsilon_{1}}{\longrightarrow} y_{p2} + x_{p1} $
& & $ \varepsilon_{1} x_{p1} y_2 $
& $ \varepsilon_{1} = (0-1)\times 10^{-4}$  molecules$^{-1}$ s$^{-1} $ \\
Deactivation of $y_{p2}$  &  $y_{p2} \stackrel{\beta_{2}}{\longrightarrow} y_2$ 
& & $ \beta_{2} y_{p2} $
& $ \beta_{2} = 0.01$ s$^{-1} $ \\
Activation of $x_3$  & $ x_3 + x_{p2} \stackrel{k_{23x}}{\longrightarrow} x_{p3} + x_{p2} $
& & $ k_{23x} x_{p2} x_3 $
& $ k_{23x} = 10^{-5}$  molecules$^{-1}$ s$^{-1} $ \\
Deactivation of $x_{p3}$  &  $x_{p3} \stackrel{\alpha_{3}}{\longrightarrow} x_3$ 
& & $ \alpha_{3} x_{p3} $
& $ \alpha_{3} = 0.01$ s$^{-1} $ \\
Activation of $y_3$  & $ y_3 + y_{p2} \stackrel{k_{23y}}{\longrightarrow} y_{p3} + y_{p2} $
& & $ k_{23y} y_{p2} y_3 $
& $ k_{23y} = 10^{-5}$  molecules$^{-1}$ s$^{-1} $ \\
Deactivation of $y_{p3}$  &  $y_{p3} \stackrel{\beta_{3}}{\longrightarrow} y_3$ 
& & $ \beta_{3} y_{p3} $
& $ \beta_{3} = 0.01$ s$^{-1} $ \\
\end{tabular}
\end{ruledtabular}
\label{table2} 
\end{table*}


\section{Results and Discussion}

\subsection{Two variable mutual information}

\begin{widetext}

\begin{figure}[!ht]
\begin{center}
\includegraphics[width=0.75\columnwidth,angle=-90]{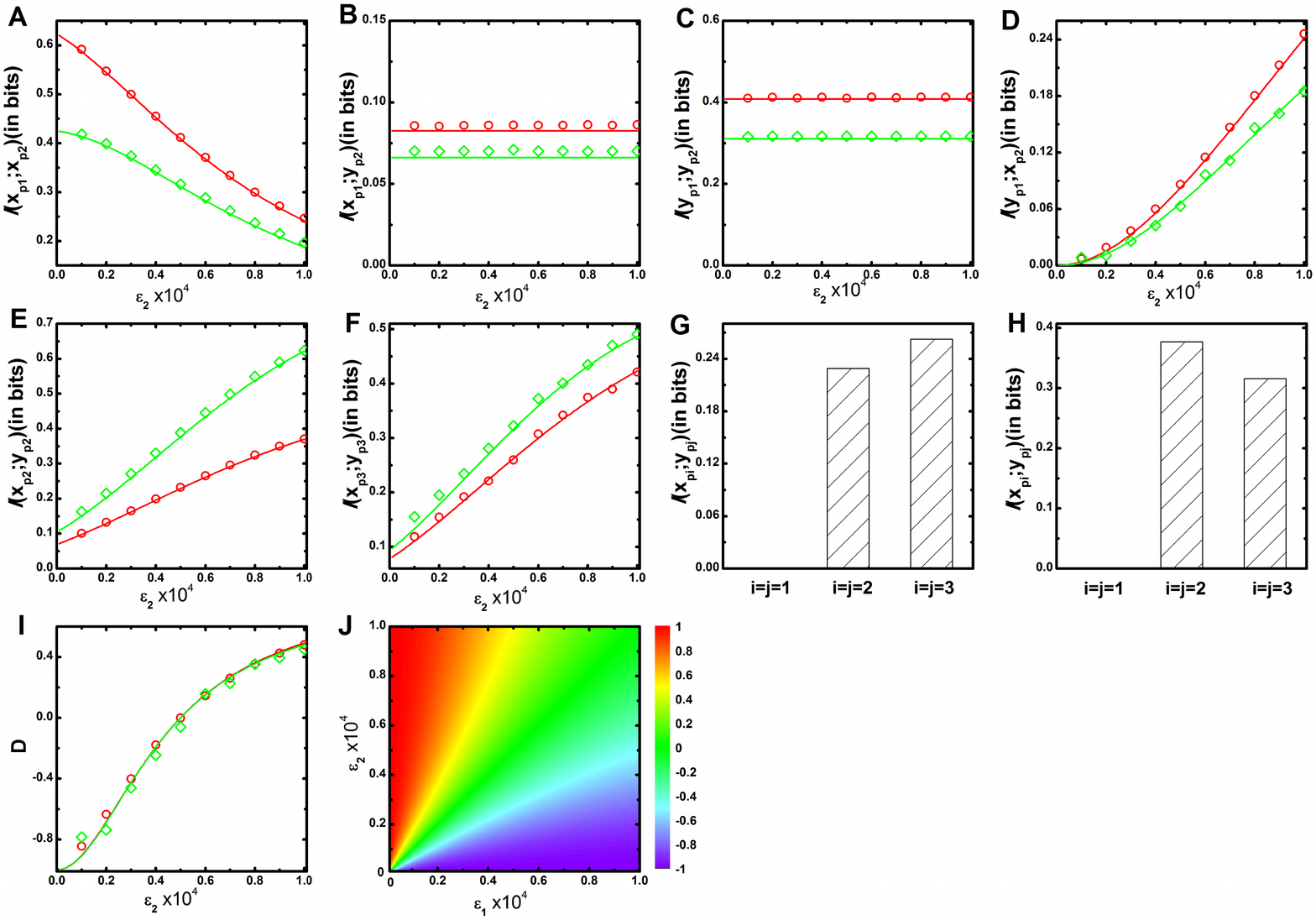}
\end{center}
\caption{(color online) 
Two variable mutual information and net information transduction as a function of
cross-talk parameter.
A, B, C, D, E and F - Two variable mutual information profiles ${\cal I} (x_{p1}; x_{p2})$, 
${\cal I} (x_{p1}; y_{p2})$, ${\cal I} (y_{p1}; y_{p2})$, ${\cal I} (y_{p1}; x_{p2})$, 
${\cal I} (x_{p2}; y_{p2})$ and ${\cal I} (x_{p3}; y_{p3})$ as a function of cross-interaction 
parameter $\varepsilon_{2}$ for a fixed value of $\varepsilon_{1} = 0.5 \times 10^{-4}$. 
In all figures, solid (with open circle) and dotted (with open diamond) lines are
generated using faster (Table~I) and slower (Table~II) relaxation rate parameters, 
respectively. 
The symbols are generated using stochastic simulation algorithm \cite{Gillespie1976}
and the lines are due to theoretical calculation.
G and H - Bar diagram of two variable mutual information of three parallel cascade kinases under an
equivalent cross-talk condition ($\varepsilon_{1} = \varepsilon_{2} = 0.5 \times 10^{-4}$) for 
faster (Table~I) and slower (Table~II) relaxation rate parameters, respectively.
I - Net information transduction $D$ as a function of cross-interaction parameter 
$\varepsilon_{2}$ 
for a fixed value of $\varepsilon_{1}= 0.5 \times 10^{-4}$. The solid (with open circle) and the dotted 
(with open diamond) lines are due to faster (Table~I) and slower (Table~II) relaxation rate 
parameters, respectively. The figure indicates data collapse for two relaxation rate parameters. 
The symbols are generated using stochastic simulation algorithm \cite{Gillespie1976} and 
the lines are obtained from theoretical calculation.
J - 2d-surface plot of net information transduction $D$ as a function of two 
cross-talk parameters $\varepsilon_{1}$ and $\varepsilon_{2}$ for faster 
(Table~I) relaxation rate parameters.
}
\label{fig2}
\end{figure}

\end{widetext}

The parameters ${\varepsilon}_2$ and ${\varepsilon}_1$ control the 
signaling channel X and Y, respectively. In Fig.~\ref{fig2}, we show the mutual 
information profile as a function of ${\varepsilon}_2$ for two different sets 
of parameters while keeping ${\varepsilon}_1$ constant. Fig. \ref{fig2}A shows 
that mutual information between $x_{p1}$ and $x_{p2}$ kinases decays 
with the increment of ${\varepsilon}_2$. Augmentation of ${\varepsilon}_2$ 
includes a competition between $x_{p1}$ and $y_{p1}$ to phosphorylate 
the $x_2$ kinase. During phosphorylation, mutual association is originated, 
and signal transduction is ensued. Thus, for the low value of ${\varepsilon}_2$, 
maximum level of mutual information is attained due to minimal phosphorylation 
competition. On the other hand, minimum level of mutual information is 
propagated at high ${\varepsilon}_2$ value due to maximum phosphorylation 
contribution of $y_{p1}$. In Fig.~\ref{fig2}B, mutual information between $x_{p1}$ and 
$y_{p2}$ is plotted, which shows a constant value as a function of ${\varepsilon}_2$. 
This happens as ${\varepsilon}_2$ has no influence in the alteration of mutual 
information. The same logic is applicable to the mutual information between 
$y_{p1}$ and $y_{p2}$ shown in Fig.~\ref{fig2}C. In Fig.~\ref{fig2}D, mutual information 
between $y_{p1}$ and $x_{p2}$ increases as a function of ${\varepsilon}_2$, 
as ${\varepsilon}_2$ is only responsible for establishing the cross-talk between 
$y_{p1}$ and $x_{p2}$. This result implies that with the enhancement of 
cross-talk the process of signal integration through $y_{p1}$ increases. 
The same profiles can be generated as a function of ${\varepsilon}_1$, 
while keeping ${\varepsilon}_2$ fixed. 
These results together indicate that ${\cal I} (x_{p1}; x_{p2})$ and ${\cal I} (y_{p1}; 
x_{p2})$ depend on ${\varepsilon}_2$, whereas ${\cal I} (x_{p1}; y_{p2})$ 
and ${\cal I} (y_{p1}; y_{p2})$ depend on ${\varepsilon}_1$. In Fig.~\ref{fig2}A-\ref{fig2}D, 
the dotted lines drawn for slower relaxation rate (see Table~II) always maintains 
a lower mutual information value compared to the solid lines drawn for faster 
relaxation rate (see Table~I). Relaxation rates of the corresponding kinases 
i.e., $x_{p1}$, $x_{p2}$, $x_{p3}$, $y_{p1}$, $y_{p2}$ and $y_{p3}$ are 
$- J_{x1x1} = ({\alpha}_1 + k_x s_x )$, $-J_{x2x2} = 
({\alpha}_2 + k_{12x} \langle x_{p1} \rangle + {\varepsilon}_2 \langle y_{p1} \rangle)$, 
$-J_{x3x3} = ({\alpha}_3 + k_{23x} \langle x_{p2} \rangle)$, $-J_{y1y1} = ({\beta}_1 + k_y s_y )$, 
$-J_{y2y2} = ({\beta}_2 + k_{12y} \langle y_{p1} \rangle + {\varepsilon}_1 \langle x_{p1} \rangle)$ 
and $-J_{y3y3} = ({\beta}_3 + k_{23y} \langle y_{p2} \rangle)$, respectively, 
where the angular bracket $\langle \cdots \rangle$ indicates the deterministic 
copy number at long time limit (see Appendix). An input signal can reliably 
flow downstream if relaxation rate (or degradation rate) of a cascade protein 
is higher than that of its upstream cascade proteins \cite{Maity2015}. For solid 
line, we consider higher degradation rate for $x_{p2}$ and $x_{p3}$ ($y_{p2}$ 
and $y_{p3}$) compared to $x_{p1}$ ($y_{p1}$). Thus faster relaxation rates 
are attained under this condition with high information propagation capacity.


Next, we quantify mutual information between two parallel kinases 
($x_{pi}$ and $y_{pj}$, with $i=j$) of the two equivalent interacting 
MAPK pathways. The inter pathway coupling is unidirectional when 
either ${\varepsilon}_1$ or ${\varepsilon}_2$ is zero but is bidirectional 
when both are non-zero. In this situation, both variables ($x_{pi}$ and 
$y_{pj}$) do not interact with each but are regulated by a common 
kinase regulon incorporating coordinated fluctuations into these variables. 
In other words, quantification of mutual information actually evaluates the 
extent of cross-correlation between these two variables. We observe zero 
mutual information value between $x_{p1}$ and $y_{p1}$, as these are 
uncorrelated. In Fig.~\ref{fig2}E, we show mutual information in between $x_{p2}$ 
and $y_{p2}$ as a function of ${\varepsilon}_2$ keeping ${\varepsilon}_1$ 
fixed. The profile shows an increasing trend as cross-talk parameter 
${\varepsilon}_2$ increases. Similarly, in Fig.~\ref{fig2}F, mutual information between 
$x_{p3}$ and $y_{p3}$ is shown with a similar trend as in Fig.~\ref{fig2}E. Interestingly, 
for faster relaxation time scale, mutual information between similar cascade 
kinases increases while moving from second ($x_{p2}$ and $y_{p2}$) to third 
($x_{p3}$ and $y_{p3}$) cascade. On the other hand, an opposite trend is 
observed for slower relaxation time scale. This characteristic trend is further 
shown in Fig.~\ref{fig2}G and \ref{fig2}H using bar diagram. These results together suggest 
that fluctuations due to faster relaxation rate transduce correlated fluctuations 
in a better way compared to the slower one. In Fig.~\ref{fig2}E, mutual information 
is high for slower relaxation rate than the faster one, as slower rate parameters 
yield high level of $x_{p2}$ and $y_{p2}$ which in turn incorporate extra 
fluctuations that help to increase mutual association. A similar result is also 
observed in Fig.~\ref{fig2}F. Identical mutual information profiles of ${\cal I} (x_{p2}; y_{p2})$ 
and ${\cal I} (x_{p3}; y_{p3})$ can be generated as function of ${\varepsilon}_1$ 
keeping ${\varepsilon}_2$ fixed.
These results suggest that both the cross-talk parameters ${\varepsilon}_1$ 
and ${\varepsilon}_2$ contribute equally to the development of an association 
between two parallel pathways.

Both the mutual information between $x_{p2}$ and $y_{p2}$, $x_{p3}$ and 
$y_{p3}$ are capable of providing a satisfactory explanation of enhancement 
of cross-talk with the increment of inter pathway interaction parameters 
(${\varepsilon}_1$ and ${\varepsilon}_2$). Under equivalent interactions 
condition (${\varepsilon}_1 = {\varepsilon}_2$), each pathway shares its 
information with other in an equal extent and is quantified not only by 
${\cal I} (x_{p2}; y_{p2})$ and ${\cal I} (x_{p3}; y_{p3})$ but also by 
${\cal I} (x_{p1}; y_{p2})$ and ${\cal I} (y_{p1}; x_{p2})$. However, characterization 
of the direction of information transduction is difficult under unequal condition 
(${\varepsilon}_1 \neq {\varepsilon}_2$). Except the equivalent condition 
(${\varepsilon}_1 = {\varepsilon}_2$) where the net information 
( ${\cal I} (y_{p1}; x_{p2} ) - {\cal I} (x_{p1}; y_{p2})$  ) flow is zero, it has a definite 
value with directionality (positive or negative value) at all other conditions. Since 
the definition of mutual information is symmetric in nature and usage of the same 
is difficult to provide directionality of information propagation, we define a 
dimensionless quantity, \textit{net information transduction} ($D$) using 
${\cal I} (x_{p1}; y_{p2})$ and ${\cal I} (y_{p1}; x_{p2})$ as
\begin{equation}
\label{eq13}
D = \frac{ 
{\cal I} (y_{p1}; x_{p2}) - {\cal I} (x_{p1}; y_{p2})
}{ 
{\cal I} (y_{p1}; x_{p2}) + {\cal I} (x_{p1}; y_{p2})
} .
\end{equation}

\noindent
The above expression implies that it is maximum ($D = 1$) when 
${\cal I} (x_{p1}; y_{p2})$ is zero, i.e., no information propagation from $x_{p1}$ 
to $y_{p2}$ (${\varepsilon}_1=0$). It is minimum ($D = -1$) 
when ${\cal I} (y_{p1}; x_{p2})$ is zero, which specifies zero information 
propagation from $y_{p1}$ to $x_{p2}$ (${\varepsilon}_2=0$). In Fig.~\ref{fig2}I, we 
show the profile of $D$ as a function of ${\varepsilon}_2$ while 
keeping ${\varepsilon}_1$ fixed, where the value of $D$ changes 
from negative to positive as ${\varepsilon}_2$ increases. It suggests that at 
low ${\varepsilon}_2$, information flowing from X to Y pathway dominates 
over the flow from Y to X. In other words, in this regime, the net information 
flow is accounted for by X $\rightarrow$ Y, leading to a negative value of 
$D$. On the other hand, at high ${\varepsilon}_2$, the direction 
of net information propagation is from Y to X due to reverse situation and 
generates a positive $D$ value. The opposite scenario can be observed 
if one generates the profile of $D$ as a function of ${\varepsilon}_1$ 
for fixed ${\varepsilon}_2$ 
In this connection, it is important to mention that both the relaxation time 
scale limits generate a similar profile of $D$. As a result, both the profiles 
of $D$ exhibit data collapse when depicted as a function of ${\varepsilon}_2$ 
for fixed ${\varepsilon}_1$ (Fig.~\ref{fig2}I) or vice versa 
This observation indicates that normalised profiles of $D$ are independent 
of relaxation time scales. In Fig.~\ref{fig2}J, we also show a 2d-surface plot of $D$ 
as a function of both ${\varepsilon}_1$ and ${\varepsilon}_2$ for faster 
relaxation time scale (Table~I). The surface plot indicates zero (or near to zero) 
value of $D$ along the diagonal region (${\varepsilon}_1 \approx {\varepsilon}_2$). 
However, the off diagonal region is positive for ${\varepsilon}_1 < {\varepsilon}_2$ 
and negative for ${\varepsilon}_1 > {\varepsilon}_2$. 


\begin{widetext}

\begin{figure}[!ht]
\begin{center}
\includegraphics[width=\columnwidth,angle=0]{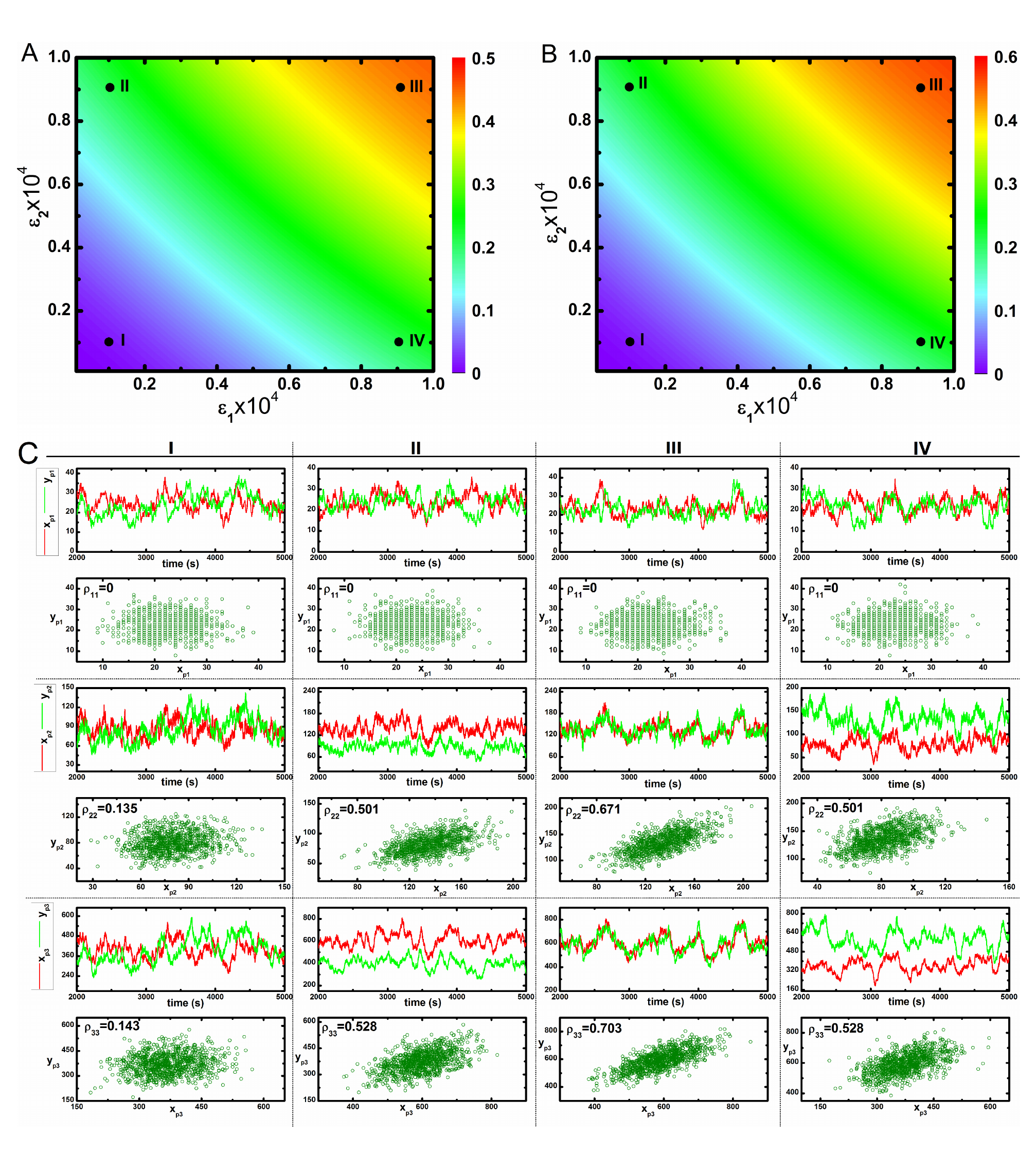}
\end{center}
\caption{(color online) 
2d-surface plots of two variable mutual information, stochastic time trajectories and
scattered plots \cite{Gillespie1976}.
A and B - 2d-surface plot of two variable mutual information ${\cal I} (x_{p2}; y_{p2})$ and 
${\cal I} (x_{p3}; y_{p3})$ as a function of two cross-talk parameters $\varepsilon_{1}$ 
and $\varepsilon_{2}$ for faster (Table~I) relaxation rate parameters. In both figures 
I, II, III and IV correspond to four different values of $\varepsilon_{1}$ and 
$\varepsilon_{2}$.
C - Stochastic time trajectories and steady state population of two parallel kinases for 
four different sets of $\varepsilon_{1}$ and $\varepsilon_{2}$. For CI, CII, CIII and CIV 
we have used 
$\varepsilon_{1} = \varepsilon_{2} = 0.1 \times 10^{-4},$ 
$\varepsilon_{1} = 0.1 \times 10^{-4}$ and 
$\varepsilon_{2} = 0.9 \times 10^{-4}$, 
$\varepsilon_{1} = \varepsilon_{2} = 0.9 \times 10^{-4}$ and 
$\varepsilon_{1} = 0.9 \times 10^{-4}$ and $\varepsilon_{2} = 0.1 \times 10^{-4}$, 
respectively. 
In each scattered plot, $\rho_{ij} (i=j)$ represents analytical value of Pearson's 
correlation coefficient. 
}
\label{fig3}
\end{figure}

\end{widetext}

\begin{widetext}

\begin{figure}[!ht]
\begin{center}
\includegraphics[width=\columnwidth,angle=0]{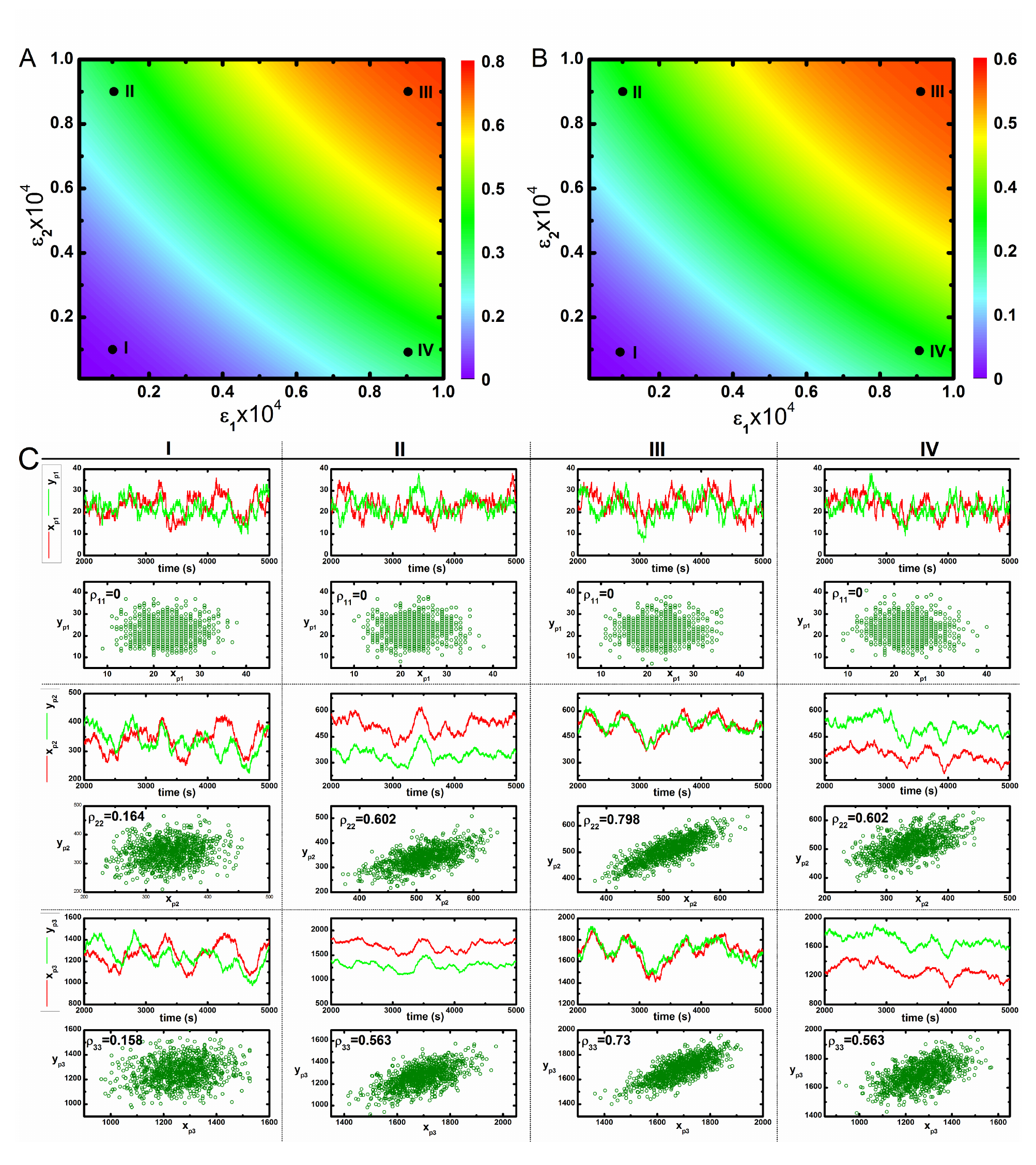}
\end{center}
\caption{(color online) 
2d-surface plots of two variable mutual information, stochastic time trajectories and
scattered plots \cite{Gillespie1976}.
A and B - 2d-surface plot of two variable mutual information ${\cal I} (x_{p2}; y_{p2})$ and 
${\cal I} (x_{p3}; y_{p3})$ as a function of two cross-talk parameters $\varepsilon_{1}$ 
and $\varepsilon_{2}$ for slower (Table~II) relaxation rate parameters. In both figures 
I, II, III and IV correspond to four different values of $\varepsilon_{1}$ and 
$\varepsilon_{2}$.
C - Stochastic time trajectories and steady state population of two parallel kinases for 
four different sets of $\varepsilon_{1}$ and $\varepsilon_{2}$. For CI, CII, CIII and CIV 
we have used 
$\varepsilon_{1} = \varepsilon_{2} = 0.1 \times 10^{-4},$ 
$\varepsilon_{1} = 0.1 \times 10^{-4}$ and 
$\varepsilon_{2} = 0.9 \times 10^{-4}$, 
$\varepsilon_{1} = \varepsilon_{2} = 0.9 \times 10^{-4}$ and 
$\varepsilon_{1} = 0.9 \times 10^{-4}$ and $\varepsilon_{2} = 0.1 \times 10^{-4}$, 
respectively. 
In each scattered plot, $\rho_{ij} (i=j)$ represents analytical value of Pearson's 
correlation coefficient. 
}
\label{fig4}
\end{figure}

\end{widetext}

In Fig.~\ref{fig3}A and \ref{fig3}B, we show two 2d-surface plots of mutual information 
between $x_{p2}$ and $y_{p2}$, $x_{p3}$ and $y_{p3}$ kinases, respectively, 
as a function of two cross-interaction parameters ${\varepsilon}_1$ and 
${\varepsilon}_2$ under faster relaxation time scale (Table~I). Both figures 
show maximum mutual information at high values of the two parameters. 
Since, ${\varepsilon}_1$ and ${\varepsilon}_2$ are equally responsible for 
developing the cross-correlation between two pathways, one can check the 
effect of maximization of mutual information by the increment of any of these 
two parameters. Although we can quantify the cross-talk with the help of two 
variable mutual information, ${\cal I} (x_{p2}; y_{p2} )$ and 
${\cal I} (x_{p3}; y_{p3})$, it is difficult to get an insight how the static and 
dynamic populations of the phosphorylated kinases are correlated. To this 
end, we have checked such correlation in Fig.~\ref{fig3}C. Dynamical correlation 
is applicable to characterize the stochastic trajectories of two variables in 
a single cell. If sufficient association between two trajectories exist, then 
correlated fluctuation is observed i.e., one trajectory closely follows the other. 
Otherwise, an uncorrelated fluctuations (trajectories do not follow each other) 
are observed in the absence of cross-talk. In Fig.~\ref{fig3}C, we show stochastic 
time series of different kinases under different conditions (four different sets 
of ${\varepsilon}_1$ and ${\varepsilon}_2$ parameters have been used - 
mentioned as I, II, III and IV in Fig.~\ref{fig3}A and \ref{fig3}B). 
These time series are generated from a single run of stochastic simulation. 
We show that stochastic time series 
exhibit correlated fluctuations at high ${\varepsilon}_1$ and ${\varepsilon}_2$ 
but are uncorrelated time series are observed at low ${\varepsilon}_1$ and 
${\varepsilon}_2$. Static correlation implies cell-to-cell population variability 
between two parallel cascade kinases. In Fig.~\ref{fig3}C, we show scattered plots 
for different kinases under different cross-interaction conditions. Here, each 
symbol generated from a single realization of stochastic trajectory represents 
copies of phosphorylated kinase at steady state. For plots with high ${\varepsilon}_1$ 
and ${\varepsilon}_2$ values, most of the symbols are disposed diagonally
in a narrow strip but for low ${\varepsilon}_1$ and ${\varepsilon}_2$, symbols 
are distributed in a much larger volume of space. 
Mutual association incorporates correlated variation in populations which is 
along the diagonal axis. An uncorrelated variation along the off-diagonal axis 
represents no such association. Therefore, cross-talk enhances correlated 
variation among populations of the kinases. These results imply that cross-talk 
not only develops association within a cell but is also capable of generating 
cell-to-cell association that assists in successful development of a significant 
robust adaptation machinery as observed in the bet-hedging program under 
diverse environmental conditions \cite{Tsimring2014}. 
In addition, similar behavior can be observed under slower relaxation time scale (Table~II)  
as shown in Fig.~\ref{fig4}. The primary difference between the nature of correlation
between $(x_{p2}, y_{p2})$ and $(x_{p3}, y_{p3})$ are visible from
Fig.~\ref{fig3} and Fig.~\ref{fig4}. In Fig.~\ref{fig3}, the correlation between
$(x_{p3}, y_{p3})$ is always higher than $(x_{p2}, y_{p2})$ for all four conditions.
On the other hand, in Fig.~\ref{fig4}, it shows an opposite trend.


\subsection{Three variable mutual information} 

In the foregoing discussion, we have shown the effect of cross-talk in terms 
of conventional two variable mutual information. However, as cross-interaction 
between two pathways develops a complex network, a comprehensive study 
of three variable mutual information provides an extra insight. In the present study, 
two types of branched pathways have been considered. One is two inputs 
($x_{p1}$ and $y_{p1}$) and one output ($x_{p2}$ or $y_{p2}$) motif where 
two input signals are integrated into a single output. The other is one input 
($x_{p1}$ or $y_{p1}$) and two outputs ($x_{p2}$ and $y_{p2}$) motif where 
the input signal is bifurcated into two output signals. In this subsection, we 
investigate the efficacy of such signal integration as well as signal bifurcation. 
Since marginal and joint distributions of all cascade proteins are considered 
as approximately Gaussian, we adopt multivariate mutual information theory
to analytically estimate three variable mutual information \cite{Barrett2015,Schneidman2003}. 
Each branched motif consists of two signal propagating channels that work 
together. It is thus interesting to investigate whether these signaling channels 
perform separately and what significant change arises in the estimation of 
three variable mutual information. The change in the magnitude of mutual 
information is defined by net synergy and is evaluated using the theory of 
partial information decomposition in terms of the difference between three 
variable mutual information and two corresponding two variable mutual
information. The value of net synergy is either positive or negative; a positive 
value indicates synergy (extra information) whereas negative value measures 
redundancy (deficit of information) \cite{Barrett2015,Schneidman2003}.


\begin{widetext}

\begin{figure}[!ht]
\begin{center}
\includegraphics[width=0.75\columnwidth,angle=-90]{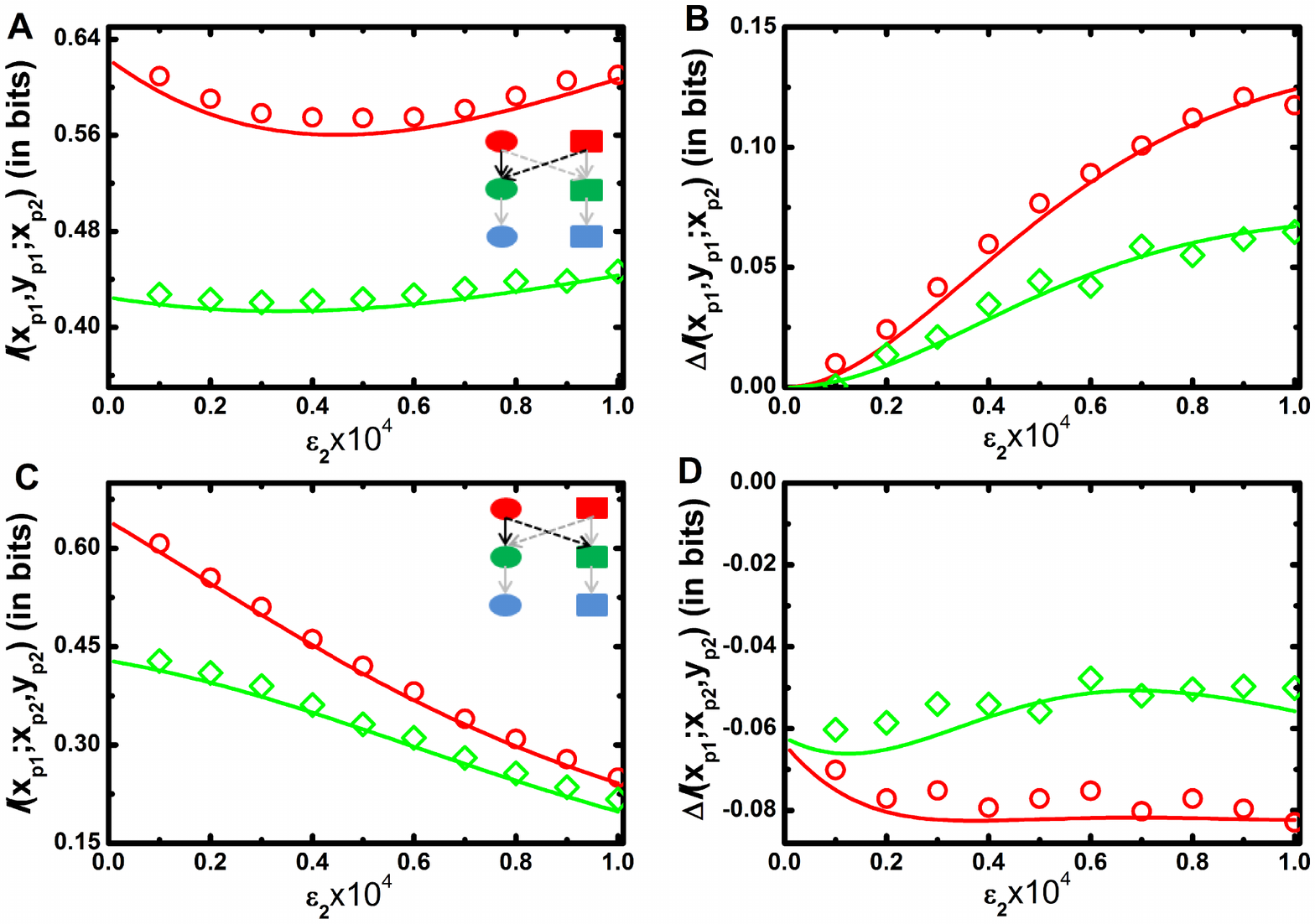}
\end{center}
\caption{ (color online)
Three variable mutual information as a function of cross-talk parameter.
A and B - Three variable mutual information ${\cal I} (x_{p1}, y_{p1}; x_{p2})$ (A) and 
net synergy $\Delta{\cal I} (x_{p1}, y_{p1}; x_{p2})$ (B) for signal integration 
motif. Schematic diagram of signal integration motif in MAPK composite network 
(see inset in A).
C and D - Three variable mutual information ${\cal I} (x_{p1}; x_{p2}, y_{p2})$ (C) and 
net synergy $\Delta{\cal I} (x_{p1}; x_{p2}, y_{p2})$ (D) for signal bifurcation 
motif. Schematic diagram of signal bifurcation motif in MAPK composite network 
(see inset in C).
All the figures are drawn as a function of cross-interaction parameter $\varepsilon_{2}$ 
for a fixed value of $\varepsilon_{1} = 0.5 \times 10^{-4}$. Here solid (with open circle) 
and dotted (with open diamond) lines are drawn for faster (Table~I) and slower (Table~II) 
relaxation rate parameters, respectively. The symbols are generated using stochastic 
simulation algorithm \cite{Gillespie1976} and the lines are obtained from theoretical 
calculation.
}
\label{fig5}
\end{figure}

\end{widetext}

In Fig.~\ref{fig5}A, we show mutual information, ${\cal I} (x_{p1}, y_{p1}; x_{p2})$ of 
two inputs and one output model as a function of ${\varepsilon}_2$ for a fixed 
value of ${\varepsilon}_1$. The profile shows a bifunctional behaviour with the 
increment of ${\varepsilon}_2$; initially it decreases up to a certain value of 
${\varepsilon}_2$, and then it increases. At low ${\varepsilon}_2$, a minimal 
amount of signal is propagated from $y_{p1}$ to $x_{p2}$. Consequently, 
the motif reduces to a single input-output motif and the motif regains its native 
form due to the significant contribution of ${\varepsilon}_2$. In Fig.~2A, we 
show that two variable mutual information between $x_{p1}$ and $x_{p2}$ 
decreases with the increment of ${\varepsilon}_2$. Similar situation arises in 
Fig.~\ref{fig5}A for low value of ${\varepsilon}_2$. On the contrary, ${\cal I} (y_{p1}; x_{p2})$ 
increases with the increment of ${\varepsilon}_2$ (Fig.~2D). Thus, two opposing 
effects work together to generate the convex profile. In Fig.~\ref{fig5}B, we plot net 
synergy of the motif as a function of ${\varepsilon}_2$ for a fixed value of 
${\varepsilon}_1$ and it is seen to increase monotonically. It is pertinent to 
mention here that for this motif, one always gets a positive net synergy value 
as a function of  ${\varepsilon}_2$. This result implies that an integrating 
signaling motif transduces more information compared to the summation of 
two isolated channels. The extra information, i.e., synergy facilitates fidelity 
of the output kinase. Intuitively, the sum of the reduction in the uncertainty 
(cross-correlation) of the output kinase contributed by each input signal is 
lower than the reduction in the uncertainty of the output provided by 
both signals together. This phenomenon implicates the aspect of integration 
of multiple signals in cellular signaling network motif as observed in 
\textit{V. harveyi} quorum-sensing circuit \cite{Waters2005,Mehta2009}

In Fig.~\ref{fig5}C, we show mutual information ${\cal I} (x_{p1}; x_{p2}, y_{p2})$ of 
one input and two outputs motif with the increment of ${\varepsilon}_2$ for a 
fixed value of ${\varepsilon}_1$. The mutual information value decreases with 
${\varepsilon}_2$ since propagation of information from $x_{p1}$ to $x_{p2}$ 
is only inhibited by the cross-interaction. However, ${\varepsilon}_2$ does not 
have any influence in information propagation from $x_{p1}$ to $y_{p2}$ and 
remains unaltered. Thus, three variable mutual information profile follows a 
decreasing trend. Fig.~\ref{fig5}D shows decreasing trend of net synergy profile as a 
function of ${\varepsilon}_2$ for a fixed value of ${\varepsilon}_1$. Importantly, 
for this motif negative values of net synergy are observed irrespective of the 
value of $\varepsilon_{2}$. This indicates redundancy in the information 
transmission in this composite motif compared to the sum of the individual one. 
Naturally, predictability about the output kinases decreases when two isolated 
signal propagation channels work together to form a bifurcated signal transduction 
motif. This result implies that although bifurcated signaling model reduces mutual 
information, it has a biological significance of the activation of multiple signaling 
channels in the presence of a single input as identified in the chemotaxis system 
of \textit{E. coli} \cite{Armitage1999}. In all figures (Fig. \ref{fig5}A-\ref{fig5}D), the solid lines are 
plotted for faster relaxation rate constants (Table~I) of $x_{p2}, x_{p3}, y_{p2}$ 
and $y_{p3}$ and the dotted lines are for slower relaxation rate constants 
(Table~II). 


\section{Conclusion}

To summarize, we have investigated evolutionarily conserved yeast MAPK signaling pathway. In our
phenomenological model, we study two parallel MAPK signaling pathways where one signaling
pathway in addition to its own activates the other pathway through cross-talk with an emphasis to
understand the change in the dynamical behavior of the system in the presence of cross-talk at the
single cell level. The model nonlinear Langevin equations have been solved under the purview of
linear noise approximation to quantify the auto variance and the covariance associated with the
different phosphorylated kinase. These quantities assist in the evaluation of mutual information (two
variable and three variable) under Gaussian channel approximation. Quantification of mutual
information has been carried out with the variation of two cross-talk parameters ${\varepsilon}_1$ 
and ${\varepsilon}_2$. The two variable mutual information shows that cross-talk establishes 
an association of signal propagation among the two pathways. To represent a better insight into 
the directionality of the net information flow, we have defined a new dimensionless 
parameter (net information transduction $D$), which varies on a scale of $-1$ to $+1$. 
Depending on the sign of $D$, we have deciphered the fidelity of one pathway 
compared to the other.

We show that cross-talk generates correlated fluctuations at the population level. A minimum and a
maximum degree of coordination are observed at low and high level of cross-talk, respectively. Our
analysis thus suggests that coordinated fluctuations are the causal effect of cross-talk in MAPK
signaling pathways. Furthermore, we have evaluated the impact of correlated association in
maintaining cell-to-cell population variability of kinases using scattered plots. At the high degree of
cross-talk, scattered plots show high correlation coefficient compared to lower level of cross talk.
These results together imply that cross-talk not only develops synchronization in a cell but also
among the cellular population. Depending on the number of inputs and outputs, we have identified
two types of signaling motifs from the composite network. In addition, quantification of three variable
mutual information allows us to calculate the net synergy associated with these two different motifs. 
The signal integration motif (two inputs and one output) reveals high fidelity, whereas the signal
bifurcation motif (one input and two outputs) shows redundancy in information propagation.

Based on the aforesaid theoretical discussion, we suggest a satisfactory explanation about the
synchronization in the outputs - a causal effect of cross-talk in parallel MAPK signaling pathways.
Nevertheless, one question obviously arises - what is the importance of such synchronization in
cellular physiology? Such functional correlation is possibly required for both the outputs to perform in
a combined way to regulate several essential downstream genes. Several experimental results on
MAPK cross-talk in \textit{S. cerevisiae} provide interesting evidence that corroborate with 
our theoretical analysis. Phosphorylated Fus3 and Kss1 are both responsible for the activation 
of transcription factor Ste12 that regulates different downstream genes \cite{Suderman2013}. 
Additionally, both activated Fus3 and Hog1 assist in arresting the cell cycle in G1 phase 
temporarily \cite{Bardwell2006,Saito2010}. Cross-talk is also highly significant for
the eukaryotic cells where the promoter of TATA binding proteins is solely controlled by MAPK
signaling pathways \cite{Gustin1998,White2010}, whereas these binding proteins are essential 
for the expression of most nuclear genes. Also, they act as a potential vehicle for 
developing coordination among the multiple disparate classes of genes. Thus, coordinated 
signaling of MAPK pathways paves the way for TATA binding proteins to establish association 
among large-scale nuclear genes. Gene regulation in \textit{S. cerevisiae} is known to be 
controlled by more than one transcription factors that bind cooperatively at
many promoter sites. This phenomenon suggests that coordinated fluctuations between the outputs of
MAPK signaling pathways are necessary to express the gene product in a controlled way. It is also
noticed that coordinated fluctuations among gene products are developed through transcriptional as
well as translational cross-talk  \cite{Dunlop2008,Komili2008,Munsky2012,Stewart2012}. 
We propose that it could be more convenient for a cell to establish a functional connection 
among all intracellular processes if the correlation is initiated in the
signaling pathway, not solely in the gene regulation stage. In fact, one interesting signature which has
been observed in different experiments is that cross-talk is prominent at low concentration level that is
manifested in diverse environmental cues \cite{Cookson2011,Tsimring2014}. Thus, in these 
situations, fluctuations in the cellular components are very high and it is very unlikely for 
cells to adopt a constructive decision for survival \cite{Junttila2008}. Our results indicate 
that such decision making program becomes easy when correlated fluctuations among the essential 
proteins are successfully implemented through the bet-hedging program \cite{Tsimring2014}.

Overall, we suggest that synchronization between MAPK signaling pathways 
is a result of cross-talk. Our analytical calculation supplemented by exact 
numerical simulation is a general approach and can be applied to other cross-talk 
pathways to quantify the strength of cross-interactions. In future, we plan to address 
the influence and physiological relevance of cross-talk in other network motifs.
Our theoretical observations 
in the present work could be verified upon the quantification of phosphorylated 
kinase protein in a single cell using flow cytometry and time lapse microscopy 
\cite{Gao2013,Schulz2007,Yaginuma2014}. These experimental approaches 
can be implemented to measure the amount of intra-cellular phosphorylated 
kinases by treating cells with external stimuli, fixing and permeabilizing cells with 
appropriate chemicals, and then staining with phospho-specific antibodies for 
different kinases. After that, one can quantify intensity of phosphorylated kinases 
in individual cells of a colony. Using these data, distribution profiles of the 
concentration of phosphorylated kinases could be developed. These quantifiable 
distribution profiles could be used to quantify the mutual information.


\begin{acknowledgments}
We thank Debi Banerjee, Sandip Kar and Jayanta Mukhopadhyay for critical 
reading and for making constructive suggestions. AKM acknowledges University 
Grants Commission for a research fellowship (UGC/776/JRF(Sc)). 
SKB acknowledges financial support from Council of Scientific and Industrial 
Research, Government of India (01(2771)/14/EMR-II)
and Bose Institute (Institutional Programme VI - Development of Systems 
Biology), Kolkata.
\end{acknowledgments}


\appendix

\section{Calculation of auto variance and covariance}

The MAPK network motif shown in Fig.~1 is explicated through stochastic 
Langevin equations. Each pathway (X or Y) is activated by the initiation of 
an extra cellular signal ($S_x$ or$ S_y$). When the first cascade kinase is 
activated, it regulates the activation of downstream kinases of the same as 
well as the parallel pathway through cross-interaction. Once activated, the 
second kinase regulates the activation of the last kinase. The activated and 
the deactivated states can be identified in terms of phosphorylated ($x_{pi}$ 
and $y_{pj}$) and dephosphorylated ($x_i$ and $y_j$) forms of each kinase 
($i,j = 1,2,3$), respectively. To construct the theoretical model of the composite 
MAPK network motif, we have considered the total population (phosphorylated 
and dephosphorylated form) of all kinases to be a constant 
($(x_i + x_{pi}) = x_{Ti} = (y_j + y_{pj}) = y_{Tj} =$ constant; $i = j$). 
Thus, for X pathway the stochastic differential equations for $x_{p1}$, 
$x_{p2}$ and $x_{p3}$ are \cite{Heinrich2002}
\begin{subequations}
\begin{eqnarray}
\frac{dx_{p1}}{dt} & = & k_x s_x (x_{T1} - x_{p1}) - {\alpha}_1 x_{p1} + \xi_1 (t), 
\\
\frac{dx_{p2}}{dt} & = & k_{12x} x_{p1} (x_{T2} - x_{p2}) + 
{\varepsilon}_2 y_{p1} (x_{T2} - x_{p2}) \nonumber \\
&& - {\alpha}_2 x_{p2} + \xi_2 (t), 
\\
\frac{dx_{p3}}{dt} & = & k_{23x} x_{p2} (x_{T3} - x_{p3}) - {\alpha}_3 x_{p3} + \xi_3 (t).
\end{eqnarray}
\end{subequations}

\noindent The first and the second terms on the right hand side of Eq.~(A.1) denote 
phosphorylation and dephosphorylation rate of the corresponding kinase. Here, 
$k_x$, $k_{12x}$ and $k_{23x}$ are activation and ${\alpha}_1$, ${\alpha}_2$ 
and ${\alpha}_3$ are deactivation rate constants of $x_{p1}$, $x_{p2}$ and $x_{p3}$, 
respectively. ${\varepsilon}_2$ is the cross-interaction parameter that controls 
signal propagation from Y to X pathway ($y_{p1} \rightarrow x_{p2}$). The 
$\xi_i$-s ($i = 1, 2, 3$) are Gaussian white noise terms with zero mean and 
finite noise strength. While writing Eq.~(A.1) we have used the conservation 
relation $x_i = x_{Ti} - x_{pi}$. Similarly, the stochastic Langevin equations 
associated with the components of the Y pathway can be written as \cite{Heinrich2002}
\begin{subequations}
\begin{eqnarray}
\frac{dy_{p1}}{dt} & = & k_y s_y (y_{T1} - y_{p1}) - {\beta}_1 y_{p1} + \eta_1 (t), \\
\frac{dy_{p2}}{dt} & = & k_{12y} y_{p1} (y_{T2} - y_{p2}) + 
{\varepsilon}_1 x_{p1} (y_{T2} - y_{p2}) \nonumber \\
&& - {\beta}_2 y_{p2} + \eta_2 (t), \\
\frac{dy_{p3}}{dt} & = & k_{23y} y_{p2} (y_{T3} - y_{p3}) - {\beta}_3 y_{p3} + \eta_3 (t).
\end{eqnarray}
\end{subequations}

\noindent In Eq.~(A.2), the first and the second terms stand for phosphorylation 
and dephosphorylation rate. Here, $k_y$, $k_{12y}$ and $k_{23y}$ are activation 
and ${\beta}_1$, ${\beta}_2$ and ${\beta}_3$ are deactivation rate constants of 
$y_{p1}$, $y_{p2}$ and $y_{p3}$, respectively. The cross-interaction parameter 
is ${\varepsilon}_1$ that controls signal transduction from X to Y pathway 
($x_{p1} \rightarrow y_{p2}$). The noise terms $\eta_i$-s ($i = 1,2,3$) are 
considered to be Gaussian white noise with zero mean and finite noise strength. 
For Y pathway, constant constraint $y_i = y_{Ti} - y_{pi}$ is also valid. The 
statistical properties of $\xi_i$-s and $\eta_j$-s ($i,j=1,2,3$) are
\begin{eqnarray*}
\langle \xi_1 \rangle = \langle \xi_2 \rangle & = & \langle \xi_3 \rangle = \langle \eta_1 \rangle = 
\langle \eta_2 \rangle = \langle \eta_3 \rangle = 0, 
\nonumber \\
\langle \xi_i(t) \xi_j(t^{\prime}) \rangle & = & \langle |\xi_i|^2 \rangle \delta_{ij} \delta (t-t^{\prime}), 
\nonumber \\
\langle \eta_i(t) \eta_j(t^{\prime}) \rangle & = & \langle |\eta_i|^2 \rangle \delta_{ij} \delta (t-t^{\prime}),
\nonumber \\
\langle \xi_i(t) \eta_j(t^{\prime}) \rangle & = & \langle |\xi_i \eta_j| \rangle \delta_{ij} \delta (t-t^{\prime}), 
\nonumber \\
\langle |\xi_1|^2 \rangle & = & k_x s_x (x_{T1} - \langle x_{p1} \rangle) + 
{\alpha}_1 \langle x_{p1} \rangle  = 2 {\alpha}_1 \langle x_{p1} \rangle,
\nonumber \\ 
\langle |\xi_2|^2 \rangle & = & (k_{12x} \langle x_{p1} \rangle (x_{T2} - \langle x_{p2} \rangle) 
\nonumber \\
&& + {\varepsilon}_2 \langle y_{p1} \rangle (x_{T2} - \langle x_{p2} \rangle) 
+ {\alpha}_2 \langle x_{p2} \rangle \nonumber \\
& = & 2 {\alpha}_2 \langle x_{p2} \rangle,
\nonumber \\
\langle |\xi_3|^2 \rangle & = & k_{23x} \langle x_{p2} \rangle (x_{T3} - \langle x_{p3} \rangle) 
+ {\alpha}_3 \langle x_{p3} \rangle \nonumber \\
& = & 2 {\alpha}_3 \langle x_{p3} \rangle,
\nonumber \\ 
\langle |\eta_1|^2 \rangle & = & k_y s_y (y_{T1} - \langle y_{p1} \rangle) 
+ {\beta}_1 \langle y_{p1} \rangle \nonumber \\
& = & 2 {\beta}_1 \langle y_{p1} \rangle,
\nonumber \\ 
\langle |\eta_2|^2 \rangle & = & k_{12y} \langle y_{p1} \rangle (y_{T2} - \langle y_{p2} \rangle) 
\nonumber \\
&& + {\varepsilon}_1 \langle x_{p1} \rangle (y_{T2} - \langle y_{p2} \rangle )
+ {\beta}_2 \langle y_{p2} \rangle \nonumber \\ 
& = & 2 {\beta}_2 \langle y_{p2} \rangle,
\nonumber \\
\langle |\eta_3|^2 \rangle & = & k_{23y} \langle y_{p2} \rangle (y_{T3} - \langle y_{p3} \rangle) 
+ {\beta}_3 \langle y_{p3} \rangle \nonumber \\
& = & 2 {\beta}_3 \langle y_{p3} \rangle,
\nonumber \\
\langle |\xi_1 \eta_1| \rangle & = & \langle |\xi_1 \eta_2| \rangle = \langle |\xi_1 \eta_3| \rangle 
= \langle |\xi_2 \eta_1| \rangle = \langle |\xi_2 \eta_2| \rangle \nonumber \\
& = & \langle |\xi_2 \eta_3| \rangle 
= \langle |\xi_3 \eta_1| \rangle = \langle |\xi_3 \eta_1| \rangle = \langle |\xi_3 \eta_3| \rangle 
\nonumber \\
& = & 0.
\end{eqnarray*}

To solve the nonlinear Eqs.~(A.1-A.2), we adopt linear noise approximation 
\cite{deRonde2012,Elf2003,Gardiner2009,Mehta2008,Paulsson2004,Paulsson2005,Thattai2004,
Kampen2011,Grima2011,Thomas2013,ThomasP2013,Maity2014,Grima2015}. 
Linearizing Eqs.~(A.1-A.2) around steady state $\delta z(t) = z(t) - \langle z \rangle$, 
where $\langle z \rangle$ is the average population of $z$ at long time limit, 
one arrives at
\begin{eqnarray}
\frac{d}{dt}
\left (
\begin{array}{c}
\delta x_{p1} \\
\delta x_{p2} \\
\delta x_{p3} \\
\delta y_{p1} \\
\delta y_{p2} \\
\delta y_{p3}
\end{array}
\right )
& = &
\left (
\begin{array}{cccccc}
  J_{x1x1}  &  J_{x1x2}  &  J_{x1x3} &  J_{x1y1}  & J_{x1y2}  & J_{x1y3}\\
  J_{x2x1}  &  J_{x2x2}  &  J_{x2x3} &  J_{x2y1}  & J_{x2y2}  & J_{x2y3} \\
  J_{x3x1}  &  J_{x3x2}  &  J_{x3x3} &  J_{x3y1}  & J_{x3y2}  & J_{x3y3} \\
  J_{y1x1}  &  J_{y1x2}  &  J_{y1x3} &  J_{y1y1}  & J_{y1y2}  & J_{y1y3} \\
  J_{y2x1}  &  J_{y2x2}  &  J_{y2x3} &  J_{y2y1}  & J_{y2y2}  & J_{y2y3} \\
  J_{y3x1}  &  J_{y3x2}  &  J_{y3x3} &  J_{y3y1}  & J_{y3y2}  & J_{y3y3}
\end{array}
\right
) 
\nonumber \\
&& \times
\left (
\begin{array}{c}
\delta x_{p1} \\
\delta x_{p2} \\
\delta x_{p3} \\
\delta y_{p1} \\
\delta y_{p2} \\
\delta y_{p3}
\end{array}
\right )
+
\left (
\begin{array}{c}
\xi_{1} \\
\xi_{2} \\
\xi_{3} \\
\eta_{1} \\
\eta_{2} \\
\eta_{3} 
\end{array}
\right ) .
\end{eqnarray}

\noindent Here
\begin{eqnarray*}  
J_{x1x1} & = & - (k_x s_x + {\alpha}_1), \nonumber \\
J_{x1x2} & = & J_{x1x3} = J_{x1y1} = J_{x1y2} = J_{x1y3} =0, \nonumber \\
J_{x2x1} & = & k_{12x} (x_{T2} - \langle x_{p2} \rangle), \nonumber \\
J_{x2x2} & = & - (k_{12x} \langle x_{p1} \rangle + {\varepsilon}_2 \langle y_{p1} \rangle + {\alpha}_2), \nonumber \\
J_{x2y1} & = & {\varepsilon}_2 (x_{T2} - \langle x_{p2} \rangle), \nonumber \\
J_{x2x3} & = & J_{x2y2} = J_{x2y3} =0, \nonumber \\ 
J_{x3x1} & = & J_{x3y1} = J_{x3y2} = J_{x3y3} =0, \nonumber \\
J_{x3x2}  & = & k_{23x} (x_{T3} - \langle x_{p3} \rangle), \nonumber \\ 
J_{x3x3} & = & -(k_{23x} \langle x_{p2} \rangle + {\alpha}_3), \nonumber \\ 
J_{y1y1} & = & - (k_y s_y + {\beta}_1), \nonumber \\
J_{y1x1} & = & J_{y1x2} = J_{y1x3} = J_{y1y2} = J_{y1y3} =0, \nonumber \\
J_{y2y1} & = & k_{12y} (y_{T2} - \langle y_{p2} \rangle), \nonumber \\
J_{y2y2} & = & - (k_{12y} \langle y_{p1} \rangle + {\varepsilon}_1 \langle x_{p1} \rangle + {\beta}_2), \nonumber \\
J_{y2x1} & = & {\varepsilon}_1 (y_{T2} - \langle y_{p2} \rangle), \nonumber \\
J_{y2y3} & = & J_{y2x2} = J_{y2x3} =0, \nonumber \\ 
J_{y3y1} & = & J_{y3x1} = J_{y3x2} = J_{y3x3} =0, \nonumber \\
J_{y3y2}  & = & k_{23y} (y_{T3} - \langle y_{p3} \rangle), \nonumber \\ 
J_{y3y3} & = & -(k_{23y} \langle y_{p2} \rangle + {\beta}_3). \nonumber \\ 
\end{eqnarray*}

\noindent 
The generalised matrix form of Eq.~(A.3) is 
\begin{equation}
\frac{d \delta \mathbf{A}}{dt} = \mathbf{J}_{A=\langle A \rangle} \delta \mathbf{A} (t) 
+ \mathbf {\Theta} (t) ,
\end{equation}

\noindent
where $\mathbf{J}$ is the Jacobian matrix evaluated at steady state. The diagonal 
elements of $\mathbf{J}$ matrix define the relaxation rate of each kinase and the 
off-diagonal elements represent the interaction rate between two different kinases 
\cite{deRonde2012,Mehta2008,Thattai2004}. Moreover, $\delta \mathbf{A}$ and 
$\mathbf{\Theta}$ are the fluctuations matrix and the noise matrix of the kinases, 
respectively. 
To calculate the different variance and covariance in the stationary state we now 
make use of the Lyapunov matrix equation \cite{Elf2003,Paulsson2004,Paulsson2005}
\begin{equation}
\mathbf{J \sigma} +\mathbf{\sigma}\mathbf{J}^T + \mathbf{D} = 0,
\end{equation}

\noindent
where $\mathbf{\sigma}$ is the covariance matrix and 
$\mathbf{D} = \langle \mathbf{\Theta}  \mathbf{\Theta}^T \rangle$ is the diffusion
matrix that depends on different noise strength. Here $\langle \cdots \rangle$ 
represents ensemble average and $T$ stands for transpose of a matrix.
Solution of Eq.~(A.5) provides the expressions of auto variance and covariance 
of the kinases
\begin{subequations}
\begin{eqnarray}
{\sigma}^2_{x_{p1}} & = & \frac{\alpha_{1} \langle x_{p1} \rangle }{J_{x1x1}} .\\
{\sigma}^2_{x_{p2}} & = & \frac{\alpha_{2} \langle x_{p2} \rangle }{J_{x2x2}} 
+ \frac{\alpha_{1} \langle x_{p1} \rangle J^2_{x2x1} }{J_{x1x1} J_{x2x2} (J_{x1x1} + J_{x2x2})} 
\nonumber \\
&& + \frac{\beta_{1} \langle y_{p1} \rangle J^2_{x2y1} }{J_{y1y1} J_{x2x2} (J_{y1y1} + J_{x2x2})} . \\
{\sigma}^2_{y_{p1}} & = & \frac{\beta_{1} \langle y_{p1} \rangle }{J_{y1y1}} .\\ 
{\sigma}^2_{y_{p2}} & = & \frac{\beta_{2} \langle y_{p2} \rangle }{J_{y2y2}} 
+ \frac{\beta_{1} \langle y_{p1} \rangle J^2_{y2y1} }{J_{y1y1} J_{y2y2} (J_{y1y1} + J_{y2y2})} 
\nonumber \\
&& + \frac{\alpha_{1} \langle x_{p1} \rangle J^2_{y2x1} }{J_{x1x1} J_{y2y2} (J_{x1x1} + J_{y2y2})} .
\end{eqnarray}
\begin{eqnarray}  
{\sigma}^2_{x_{p1} x_{p2}} & = & {\sigma}^2_{x_{p2} x_{p1}} = \frac{\alpha_{1} \langle x_{p1} \rangle J_{x2x1} J_{x2x2}}{J_{x1x1} J_{x2x2} (J_{x1x1} + J_{x2x2})} .\\
{\sigma}^2_{x_{p1} y_{p2}} & = & {\sigma}^2_{y_{p2} x_{p1}} = \frac{\alpha_{1} \langle x_{p1} \rangle J_{y2x1} J_{y2y2}}{J_{x1x1} J_{y2y2} (J_{x1x1} + J_{y2y2})} .\\
{\sigma}^2_{y_{p1} y_{p2}} & = & {\sigma}^2_{y_{p2} y_{p1}} = \frac{\beta_{1} \langle y_{p1} \rangle J_{y2y1} J_{y2y2}}{J_{y1y1} J_{y2y2} (J_{y1y1} + J_{y2y2})} .\\
{\sigma}^2_{y_{p1} x_{p2}} & = & {\sigma}^2_{x_{p2} y_{p1}} = \frac{\beta_{1} \langle y_{p1} \rangle J_{x2y1} J_{x2x2}}{J_{y1y1} J_{x2x2} (J_{y1y1} + J_{x2x2})} .\\
{\sigma}^2_{x_{p1} y_{p1}} & = & {\sigma}^2_{y_{p1} x_{p1}} = 0 .
\end{eqnarray}
\begin{widetext}
\begin{eqnarray}
{\sigma}^2_{x_{p3}} & = & \frac{\alpha_{3} \langle x_{p3} \rangle }{J_{x3x3}} 
+ \frac{\alpha_{2} \langle x_{p2} \rangle J^2_{x3x2} }{J_{x2x2} J_{x3x3} (J_{x2x2} + J_{x3x3})} 
+ \frac{\alpha_{1} \langle x_{p1} \rangle J^2_{x2x1} J^2_{x3x2} (J_{x1x1} + J_{x2x2} + J_{x3x3}) }{J_{x1x1} J_{x2x2} J_{x3x3} (J_{x1x1} + J_{x2x2}) (J_{x1x1} + J_{x3x3}) (J_{x2x2} + J_{x3x3})} 
\nonumber \\
&& + \frac{\beta_{1} \langle y_{p1} \rangle J^2_{x2y1} J^2_{x3x2} (J_{y1y1} + J_{x2x2} + J_{x3x3}) }{J_{y1y1} J_{x2x2} J_{x3x3} (J_{y1y1} + J_{x2x2}) (J_{y1y1} + J_{x3x3}) (J_{x2x2} + J_{x3x3}) } . \\
{\sigma}^2_{y_{p3}} & = & \frac{\beta_{3} \langle y_{p3} \rangle }{J_{y3y3}} 
+ \frac{\beta_{2} \langle y_{p2} \rangle J^2_{y3y2} }{J_{y2y2} J_{y3y3} (J_{y2y2} + J_{y3y3})} 
+ \frac{\beta_{1} \langle y_{p1} \rangle J^2_{y2y1} J^2_{y3y2} (J_{y1y1} + J_{y2y2} + J_{y3y3}) }
{J_{y1y1} J_{y2y2} J_{y3y3} (J_{y1y1} + J_{y2y2}) (J_{y1y1} + J_{y3y3}) (J_{y2y2} + J_{y3y3})} 
\nonumber \\ 
&& + \frac{\alpha_{1} \langle x_{p1} \rangle J^2_{y2x1} J^2_{y3y2} (J_{x1x1} + J_{y2y2} + J_{y3y3}) }{J_{x1x1} J_{y2y2} J_{y3y3} (J_{x1x1} + J_{y2y2}) (J_{x1x1} + J_{y3y3}) (J_{y2y2} + J_{y3y3}) } . \\
{\sigma}^2_{x_{p2} y_{p2}} & = & {\sigma}^2_{y_{p2} x_{p2}} \nonumber \\
& = & 
\frac{\alpha_{1} \langle x_{p1} \rangle J_{x2x1} J_{y2x1} (J_{x2x2} + J_{y2y2} + 2 J_{x1x1}) }
{J_{x1x1} (J_{x2x2} + J_{y2y2}) (J_{x2x2} + J_{x1x1}) (J_{y2y2} + J_{x1x1}) } 
+ \frac{\beta_{1} \langle y_{p1} \rangle J_{x2y1} J_{y2y1} (J_{x2x2} + J_{y2y2} + 2 J_{y1y1}) }
{J_{y1y1} (J_{x2x2} + J_{y2y2}) (J_{x2x2} + J_{y1y1}) (J_{y2y2} + J_{y1y1}) } . \nonumber \\
&& \\
{\sigma}^2_{x_{p3} y_{p3}} & = & {\sigma}^2_{y_{p3} x_{p3}} \nonumber \\ 
& = & \frac{
\alpha_{1} \langle x_{p1} \rangle J_{x3x2} J_{y3y2} J_{x2x1} J_{y2x1} C_{1}
}{
\begin{array}{c}
J_{x1x1} (J_{x2x2} + J_{y2y2}) (J_{x2x2} + J_{y3y3}) (J_{x3x3} + J_{y2y2}) (J_{x3x3} + J_{y3y3}) \\\times (J_{x2x2} + J_{x1x1}) (J_{x3x3} + J_{x1x1}) (J_{y2y2} + J_{x1x1}) (J_{y3y3} + J_{x1x1})
\end{array}
} \nonumber \\ 
&& + \frac{
\beta_{1} \langle y_{p1} \rangle J_{x3x2} J_{y3y2} J_{x2y1} J_{y2y1} C_{2}
}{
\begin{array}{c}
J_{y1y1} (J_{x2x2} + J_{y2y2}) (J_{x2x2} + J_{y3y3}) (J_{x3x3} + J_{y2y2}) (J_{x3x3} + J_{y3y3}) \\\times (J_{x2x2} + J_{y1y1}) (J_{x3x3} + J_{y1y1}) (J_{y2y2} + J_{y1y1}) (J_{y3y3} + J_{y1y1})
\end{array}
} .
\end{eqnarray}

\noindent Here,
\begin{eqnarray}
C_{1} & = & (J_{x2x2} + J_{y2y2}) (J_{x2x2} + J_{y3y3}) (J_{x3x3} + J_{y2y2}) (J_{x3x3} + J_{y3y3}) 
+ 2 J_{x1x1} ((J_{x3x3} + J_{y2y2}) \nonumber \\ 
&& (J_{x3x3} + J_{y3y3}) (J_{y2y2} + J_{y3y3}) + J^2_{x2x2} (J_{x3x3} + J_{y2y2} + J_{y3y3}) + J_{x2x2} (J_{x3x3} + J_{y2y2} + J_{y3y3})^2) \nonumber \\ 
&& + 2 J^2_{x1x1} (J_{x2x2} + J_{x3x3} + J_{y2y2} + J_{y3y3})^2 + 2  J^3_{x1x1} (J_{x2x2} + J_{x3x3} + J_{y2y2} + J_{y3y3}) ,\nonumber \\
C_{2} & = & (J_{x2x2} + J_{y2y2}) (J_{x2x2} + J_{y3y3}) (J_{x3x3} + J_{y2y2}) (J_{x3x3} + J_{y3y3}) 
+ 2 J_{y1y1} ((J_{x3x3} + J_{y2y2}) \nonumber \\ 
&& (J_{x3x3} + J_{y3y3}) (J_{y2y2} + J_{y3y3}) + J^2_{x2x2} (J_{x3x3} + J_{y2y2} + J_{y3y3}) + J_{x2x2} (J_{x3x3} + J_{y2y2} + J_{y3y3})^2) \nonumber \\ 
&& + 2 J^2_{y1y1} (J_{x2x2} + J_{x3x3} + J_{y2y2} + J_{y3y3})^2 + 2  J^3_{y1y1} (J_{x2x2} + J_{x3x3} + J_{y2y2} + J_{y3y3}) .\nonumber 
\end{eqnarray}
\end{widetext}
\end{subequations}

In our calculation, we use the analytical expressions of auto variance and 
covariance for evaluating the value of mutual information and correlation 
coefficient.


\end{document}